\documentclass{aa}  

\usepackage{natbib}
\usepackage{graphicx}
\usepackage{amsmath}
\usepackage{mhchem}
\usepackage{xcolor}
\usepackage{ulem} 
\usepackage{caption}
\usepackage{float}
\usepackage[colorlinks=true, allcolors=blue]{hyperref}
\usepackage{soul} 
\usepackage{tabu}
\usepackage{xcolor}

\definecolor{darkorange}{rgb}{1.0, 0.55, 0.0}

\definecolor{darkgreen}{rgb}{0.5, .75, 0.}

\usepackage{makecell}

\usepackage{subcaption}         
\usepackage{lscape}             
\usepackage{placeins}           

\usepackage{txfonts}
\begin{document}

   \title{Predicting the detectability of sulphur-bearing molecules in the solid phase with simulated spectra of JWST instruments}
    
   \author{A. Taillard\inst{1},
           R. Martín-Doménech\inst{1}, H. Carrascosa\inst{1}, J.~A. Noble\inst{2}, G.~M. Muñoz Caro\inst{1}, E. Dartois\inst{3}, D. Navarro-Almaida\inst{1} , B. Escribano\inst{1}, \'A. S\'anchez-Monge\inst{4,5}, A. Fuente\inst{1}
          }

  \institute{Centro de Astrobiología (CAB), CSIC-INTA, Ctra. de Ajalvir, km 4, Torrejón de Ardoz, 28850 Madrid, Spain
              \email{ataillard@cab.inta-csic.es}
         \and
         Physique des Interactions Ioniques et Mol\'{e}culaires, CNRS, Aix Marseille Univ., 13397 Marseille, France
         \and
         Institut des Sciences Mol\'eculaires d’Orsay, CNRS, Univ. Paris-Saclay, 91405 Orsay, France
         \and
         Institut de Ci\`encies de l'Espai (ICE, CSIC), Campus UAB, Carrer de Can Magrans s/n, 08193, Bellaterra (Barcelona), Spain
         \and
         Institut d'Estudis Espacials de Catalunya (IEEC), 08860 Castelldefels (Barcelona), Spain
         }

   \date{\today}

 
  \abstract
   {
    To date, gas phase observations of sulphur in dense interstellar environments have only constrained the molecular carriers of $\sim$~1~\% of its predicted cosmic abundance. An additional $\sim$~5~\% is known to be locked up in molecular solids in dense clouds, leaving the main reservoir of depleted sulphur in the solid phase yet to be identified. Overall, OCS is the only S-bearing molecule unambiguously detected in interstellar ices thus far with infrared telescopes, although an absorption feature of SO$_2$ has been plausibly identified at 7.5 $\mu$m. The spectral resolution and sensitivity of the \textit{James Webb} Space Telescope (JWST) could make a substantial difference in detecting part of this missing sulphur. The wavelength coverage of the JWST includes vibrational absorption features of the S-carriers H$_2$S, OCS, SO$_2$, CS$_2$, SO, CS, and S$_8$ are found. The aim of this study is to determine whether these molecules may be viable candidates for detection.

    We carried out new laboratory measurements of the IR absorption spectra of CS$_2$ and S$_8$ to update the IR band strength of the most intense CS$_2$ absorption feature at 6.8~$\mu$m, as well as to determine that of S$_8$ at 20.3~$\mu$m for the first time. These data, along with values previously reported in the literature for H$_2$S, OCS, and SO$_2$, allow us to evaluate which S-bearing species could be potentially detected with JWST in interstellar ices. 
    Taking the literature abundances of the major ice species determined by previous IR observations towards starless cores, low-mass young stellar objects (LYSOs) and massive young stellar objects
(MYSOs), we generated simulated IR spectra using the characteristics of the instruments on the JWST. Thus, we have been able to establish a case study for three stages of the star formation process. 

    These spectra were simulated using a tool that produces synthetic ice spectra, with the aim of studying the feasibility of detecting S-bearing species with the JWST by artificially adding S-bearing molecules to the simulated spectra.
    We conclude that the detection of S-bearing molecules remains challenging due to a variety of parameters; principally, the overlap of absorption features with those of other species and the mixing of molecular species in the ice impacting the profile and central position of the targeted bands. Despite these obstacles, the detection of H$_2$S in dense clouds -- and potentially SO$_2$ in LYSOs and MYSOs -- should be possible in regions with favourable physical and chemical conditions, but not necessarily in the same region. In contrast, the large allotrope S$_8$ would remain undetected even in the unrealistic case that all the available sulphur atoms were involved in its formation. Although the sensitivity of JWST is insufficient to determine the sulphur budget in the solid state, the detection of (or setting of significant upper limits on the abundance of ) an additional icy sulphur compound (H$_2$S, SO$_2$) would enable us to validate a state-of-the-art approach in our knowledge of sulphur chemistry, offering a unique opportunity to make comparisons against future developments. 
    
    }


   \titlerunning{Searching for S in ISM ices with synthetic IR spectra}
   \authorrunning{Taillard et al.}

   \maketitle
%

\section{Introduction}\label{intro}

With the launch of the\textit{ James Webb} Space Telescope (JWST), the combination of high sensitivity and spectral resolution offered by its spectroscopic instruments has afforded astrochemists the opportunity to search for new species in the gas phases (and, especially, the solid phases). The goal such studies would be to determine the missing link between certain molecules detected in the gas phase and their solid phase formation. At present, the main reservoirs of sulphur in the ISM remain unknown despite the fact that sulphur plays an important role in the chemistry of star forming regions. 
Atomic S regulates the gas ionisation fraction in dense molecular clouds because its ionisation potential of slightly over 10~eV lies below that of hydrogen; thus, it is the main atomic electron donor in the $\sim$ 3.7--7 visual extinction (A$_V$) magnitude range \citep{2023A&A...670A.114F}.
The gas ionisation fraction parameter controls the coupling of the gas with the magnetic field that plays a crucial role in cloud collapse, triggering the formation of protostars \citep{2013A&A...560A.114P,2016MNRAS.460.2050Z}. 
An accurate knowledge of the amount of atomic sulphur (including S$^{+}$) in the translucent part of molecular clouds is therefore mandatory to progress in our understanding of the cloud collapse process.
S-bearing molecules are also often used as chemical clocks during the star formation process: due to the protostellar heating, solid-phase species sublimate, releasing S-bearing molecules into the gas phase that further react and form new species, which can be observed in radio astronomy; for instance, SO$_2$ or H$_2$S \citep{1997ApJ...481..396C,1998A&A...338..713H,2004A&A...422..159W,2011A&A...529A.112W}. 
Shocks from protostellar outflows can be traced by sulphur compounds such as CS, SO, and SO$_2$, whose abundances are enhanced by shock-induced desorption and excitation \citep{2001A&A...370.1017V}.

The gas-phase S-bearing molecules observed in prestellar cores (mainly SO, SO$_2$, CS, HCS$^+$, H$_2$CS, C$_2$S, and C$_3$S) only account for a very small fraction ($<$ 1\%) of the cosmic abundance of sulphur \citep{2013ChRv..113.8710A,2018MNRAS.478.5514V}. Most of the sulphur in the dense ISM is thus expected to be locked in the solid phase, either in the (semi-)refractory dust grains or their molecular icy mantles \citep{1990A&A...231..466M,1999MNRAS.306..691R,vidal_reservoir_2017,2018MNRAS.476.4949D,2019ApJ...885..114K}. So far, the search for S-bearing molecules in ISM ices has been inconclusive. 
At one point, OCS was proposed as the carrier of the 4.9~$\mu$m feature in the IR spectrum observed towards W33A using the IRTF \citep{1995ApJ...449..674P} and towards embedded objects in the Taurus dark cloud using UKIRT \citep{1997ApJ...479..839P}. These authors attributed the absorption feature to the C-S stretching band of OCS based on comparison with laboratory experiments.
In later studies, detections or upper limits on OCS were reported for 21 of the 23 objects in an ISO survey of a variety of object types by \citet{2004ApJS..151...35G}, and towards a class 0-I source with AKARI \citep{2012A&A...538A..57A}. In a similar survey of 23 MYSO observed with the IRTF, 20 detections and three upper-limits on OCS were also reported in \citet{2022ApJ...941...32B}.
Most recently, \citet{2023NatAs...7..431M} reported detections of the OCS absorption feature at 4.9~$\mu$m in JWST spectra of lines of sight towards two highly extincted field stars behind the Chamaeleon I cloud, although the surrounding spectral region also contained a lot of gas phase emission lines from the stellar photospheres, which complicates the analysis. 
The $\nu_3$ vibration mode of SO$_2$ was first associated with an absorption feature around 7.5 $\mu$m towards two YSOs in \citep{Boogert_1997}, W33A, and NGC 7538 IRS 1. In \citep{2023NatAs...7..431M}, the same feature was also tentatively assigned to SO$_2$ towards the previously mentioned Chameleon I background stars.
It has also been possibly identified towards a low-mass class 0 object and a high mass protostar by \citet{Rocha2024} based on its contribution to the 7.7 $\mu$m feature, mainly dominated by CH$_4$.
Finally, a recent study by \citet{Slavicinska_2024} has tentatively attributed the 5.3 um band in two LYSOs to a combination mode (NH$_4^+$ $\nu_4$ mode and SH$^-$ libration mode) of the ammonium salt NH$_4$SH. 
These three species remain to this day the only S-carriers (possibly) identified in ISM ices.

Comets are icy bodies resulting from star formation, with the cometary material thought to have been assembled through the entire process and thermally processed by the newly born star. The cometary ices are indirectly traced from the outgassing of their comae, making their ice composition difficult to trace back to the disk or even earlier phases. 
Rosina in situ observations of the coma of comet 67P/Churyumov-Gersimenko (hereafter, 67P) \citep{2016MNRAS.462S.253C} revealed a large fraction of neutral atomic sulphur (S) was present ($\sim$ 0.48\% abundance relative to water) and that the most abundant volatile S-bearing molecule in this comet is H$_2$S ($\sim$ 1.06\% abundance relative to water), followed by SO$_2$ ($\sim$ 0.12\% abundance relative to water), SO (0.07\% abundance relative to water), and OCS (0.04\% abundance relative to water). These molecules might be favoured in term of low temperature chemistry and were probably released from the subsurface ice of the cometary nucleus by thermal desorption \citep{2017MNRAS.469S.685C,2018MNRAS.476.4949D}.
Therefore, these S-carriers make good candidates to be searched for in colder environments prior to the formation of the proto-star with a view to comparing their abundances to those observed in comets, with a particular emphasis on H$_2$S. 
However, the detection of H$_2$S in interstellar ices remains elusive, with only upper limits reported thus far in dense clouds \citep{1991MNRAS.249..172S,2023NatAs...7..431M} and YSOs \citep{1985A&A...146L...6G,1991MNRAS.251P..24G,2011A&A...536A..91J}, but still within $\lesssim$ 1\% with respect to water, which is consistent with the observed cometary abundances.
\citet{calmonte_sulphur-bearing_2016} inferred the presence of S$_3^+$ and S$_4^+$ in the coma of comet 67P that can be attributed to S$_3$ and
S$_4$ molecules, or, alternatively, to fragments of larger S-species (Carrascosa et al. in prep.); this is similar to HCN origins as a product of polymer dissociation \citep{Fray_2004}. 

\citet{2017ApJ...851L..49F} detected gas phase HS$_2$ towards the Horsehead nebula, which is thus far the only molecule containing a S-S bond observed in the ISM. It suggests the existence of longer sulphur chains, as experiments show HS$_2$ is a product of the UV irradiation of H$_2$S:H$_2$O ices alongside S-chains (S$_x$, x $\geq$ 2) \citep{jimenez-escobar_sulfur_2011}. However, the detection of large allotropes (S$_x$, x $\geq$ 3) in the ISM has not yet been reported.

The paucity of detected S-bearing molecules in interstellar ice challenges the results from astrochemical models and observations of Solar System comets with a lack of observational constraints in the solid phase. 
Current chemical models struggle when accounting for sulphur chemistry, as was highlighted and partially corrected in \citet{vidal_reservoir_2017}. Subsequently, 
\citet{2019A&A...624A.108L} performed an in-depth revision of the sulphur chemistry in the dense ISM based on their own model, namely: a modified version of the Ohio State University (OSU) gas-grain astrochemical model that is time-dependent and single-point) \citep{2008ApJ...682..283G}. The authors added numerous photodesorption channels based on \cite{2009ApJ...690.1497H}, but completely disabled chemical desorption. Moreover, they assumed enhanced accretion of cations (C$^+$, S$^+$) following \citet{1999MNRAS.306..691R}. The overall modifications helped in increasing the sulphur depletion and favouring surface reactions involving S-bearing species.
Their model predicted that a significant fraction of the volatile sulphur is present in ice mantles ($>$ 99\% of sulphur depleted on grains in their dense model after 10$^6$ years), but the exact S-carriers are strongly dependent on the initial conditions and chemical age. For cold and evolved ($\sim$ 1 Myr) dense cores, sulphur would mainly be in the form of solid SO, solid OCS, and solid CS. Their predicted abundances are consistent with gas-phase observations in various environments (TMC-1, L1314N, Orion) and upper-limits of solid OCS in embedded protostars \citep[considering the solid OCS vs the solid CO ratio from][with the best match being 5\% observed in W33A]{1997ApJ...479..839P}.
Other chemical models tried to constrain the sulphur chemistry with other physical constraints; for instance, in \citet{navarro-almaida_gas_2020}, where the authors showed that small variations of grain surface parameters, such as a diffusion-to-binding energy ratio, different efficiencies for chemical desorption, and/or an earlier chemical age would lead to different ice compositions, with H$_2$S being the most abundant component of the S-carriers, as atoms and SH have a higher mobility. The formation of H$_2$S and OCS remains very sensitive to the assumed grain parameters. 
Unfortunately, due to the non-detection of other S-bearing species in the ices (e.g. SO or CS), the comparison between models and observations is very limited, with a clear lack of constraints.

Based on ISM and cometary observations, laboratory experiments, and astrochemical models, we can identify several S-bearing species to be searched for in ice mantles in the different stages of star formation: OCS, H$_2$S, SO$_2$, SO, and CS, with the possible presence of other ice species such as H$_2$S$_2$, CS$_2$, and S allotropes. It is of paramount importance to add constraints on these abundances to understand the sulphur chemistry. 
Therefore, we need observations with better sensitivity and better spectral resolution, with our best option being JWST NIRSpec and MIRI MRS, to finally confirm the presence or absence of these S-bearing candidates that could account for a significant fraction of the missing sulphur in the dense ISM.

In this work, we use the Synthetic Ice Spectrum Generator (SynthIceSpec, Taillard et al, in prep) to evaluate the detectability of potential S-carriers in ice mantles present in star-forming regions at different evolutionary stages. The tool is briefly presented in a subsequent section.
Section~\ref{lab} reviews the IR band strength values corresponding to S-bearing molecules reported in the literature and presents new laboratory experiments used to estimate the S$_8$ IR band strength, thereby updating the CS$_2$ value.
In Sect.~\ref{SIS} we explore the potential overlapping of sulphur bands by other ice species, using synthetic ice spectra, based on laboratory data, modelling, and past observations. A simple approach with the main ice species shows the difficulties of detecting the S-carrier bands. 
Then in Sect.~\ref{detectability} we present nine synthetic spectra covering the different ice compositions found in dense cores, LYSOs, and MYSOs, using data compiled by \citet{2015ARA&A..53..541B}. We derived, for each ice composition, the threshold of detection; namely, the column density required (w.r.t. that of water) to detect the S-bearing molecules considered here in our synthetic spectra. For each spectrum, we checked the detectability in three different constant continua to represent a bright source, a faint source, and a very faint source (respectively 1, 0.1, and 0.04 mJy), along with considering the impact this has on the determination of a detection threshold. 
In Sect.~\ref{discussion}, we discuss the different environments and chemistry needed to ensure that sufficient S-bearing abundances would allow us to reach the thresholds of detection that we have determined.
Overall, the goal of this study is to identify the feasibility of the observations of S-bearing molecules that have not yet been detected in the solid phase at different stages of star formation, taking into account the JWST spectral resolving power.

\section{IR band strengths of S-bearing molecules}\label{lab} 

\begin{table*}[!h] 
\centering
\caption{Spectral data}
\begin{tabular}{|lrrrrr|r|}
\hline

  Molecule & \makecell{Wavenumber \\ (cm$^{-1}$)} & \makecell{Wavelength \\($\mu$m)} & \makecell{Band strength \\ (cm/molecule)} & Modes        & Reference\\
\hline
  H$_2$S     & 2547.0 & 3.94          & 1.69 $\times$ 10$^{-17}$    & S-H stretch  & \citet{2022ApJ...931L...4Y} \\
  CS         & 1270.0 & 7.87          & -                           & C-S sym. stretch  & \citet{bohn_1992} \\ 
  CS$_2$     & 1502.0 & 6.66          & 1.06 $\times$ 10$^{-16}$    & C-S asym. stretch  & This work\\ 
  CS$_2$     & 2145.0 & 4.66          & 2.09 $\times$ 10$^{-18}$    & C-S combination  & This work\\ 
  OCS        & 514.0  & 19.46         & 1.80 $\times$ 10$^{-18}$    & C-S bending  & \text{\citet{1993ApJS...86..713H}} \\
  OCS        & 2025.0 & 4.94          & 1.50 $\times$ 10$^{-16}$    & C-O sym. stretch  & \text{\citet{1993ApJS...86..713H}} \\
  SO$_2$     & 1323.0 & 7.49          & 4.2 $\times$ 10$^{-17}$    & S-O asym. stretch  & \text{\citet{2022ApJ...931L...4Y}} \\
  SO$_2$     & 1149.0 & 8.70          & 2.20 $\times$ 10$^{-18}$    & S-O sym. stretch  & \text{\citet{2008P&SS...56.1300G}}\\
  SO         & 1130.0 & 8.85          & 2.20 $\times$ 10$^{-18}$    & S-O stretch  & \text{\citet{2008P&SS...56.1300G}}\\
  S$_3$      & 680.0  & 14.71         & $\sim$ 2.50 $\times$ 10$^{-17}$    & S-S stretch & Estimated value (see the text)\\
  S$_4$      & 665.0  & 15.04         & $\sim$ 5.00 $\times$ 10$^{-17}$    & S-S stretch & Estimated value (see the text)\\
  S$_8$      & 471.0  & 21.23         & 1.50 $\times$ 10$^{-19}$    & S-S stretch  & This work$^a$\\
\hline\end{tabular}
\tablefoot{Spectral data for some sulphur species reported from literature derived from deposited species or after irradiation, as well as our new measurements for CS$_2$ and S$_8$. $^a$ spectrum measured at room temperature; all other species were measured at cryogenic temperatures ranging from 5.5 to 16 K.}\label{tab:spectral_data_S}
\end{table*}

The IR band strengths are fundamental parameters to predict and interpret observed IR spectra along the lines of sight containing solid phase material, since they link the integrated absorbance of the different spectral features to the column density of the corresponding species through the equation
\begin{equation}
N=\frac{1}{A}\int_{band}{\tau_{\nu} \ d\nu}, 
\label{eqn}
\end{equation}
where $N$ is the species' column density (i.e. the number of molecules per surface area), $\tau_{\nu}$ is the optical depth of the absorption band ($\approx$ 2.3 times the absorbance), and $A$ is the corresponding integrated band strength of the vibrational mode for the considered solid species.

The IR band strengths for most of the S-bearing species relevant to ice and dust grain observations can be found in the literature (Table \ref{tab:spectral_data_S}). 
However, the IR band strengths of sulphur allotropes have, to our knowledge, not been previously reported.
Section \ref{S8} presents a rough experimental estimation of the 21.35 $\mu$m S$_8$ IR band strength, the most stable sulphur allotrope.

In addition, the band strength value commonly used in the literature for the 6.65 $\mu$m CS$_2$ IR feature 
actually corresponds to the value derived from the IR spectrum of this species in the gas phase \citep{1964AcSpe..20..771P}. 
A band strength value in the ice was reported in \citet{1964JChPh..40..309Y}, but only for a pure CS$_2$ ice in crystalline form at 77 K. 
Section \ref{CS2} presents new laboratory experiments aimed at calculating the band strength of this feature for CS$_2$ ice in amorphous form.

\subsection{Laboratory measurements of the S$_8$ IR feature band strength}\label{S8}

\begin{figure}
    \centering
    \includegraphics[width=0.99\linewidth]{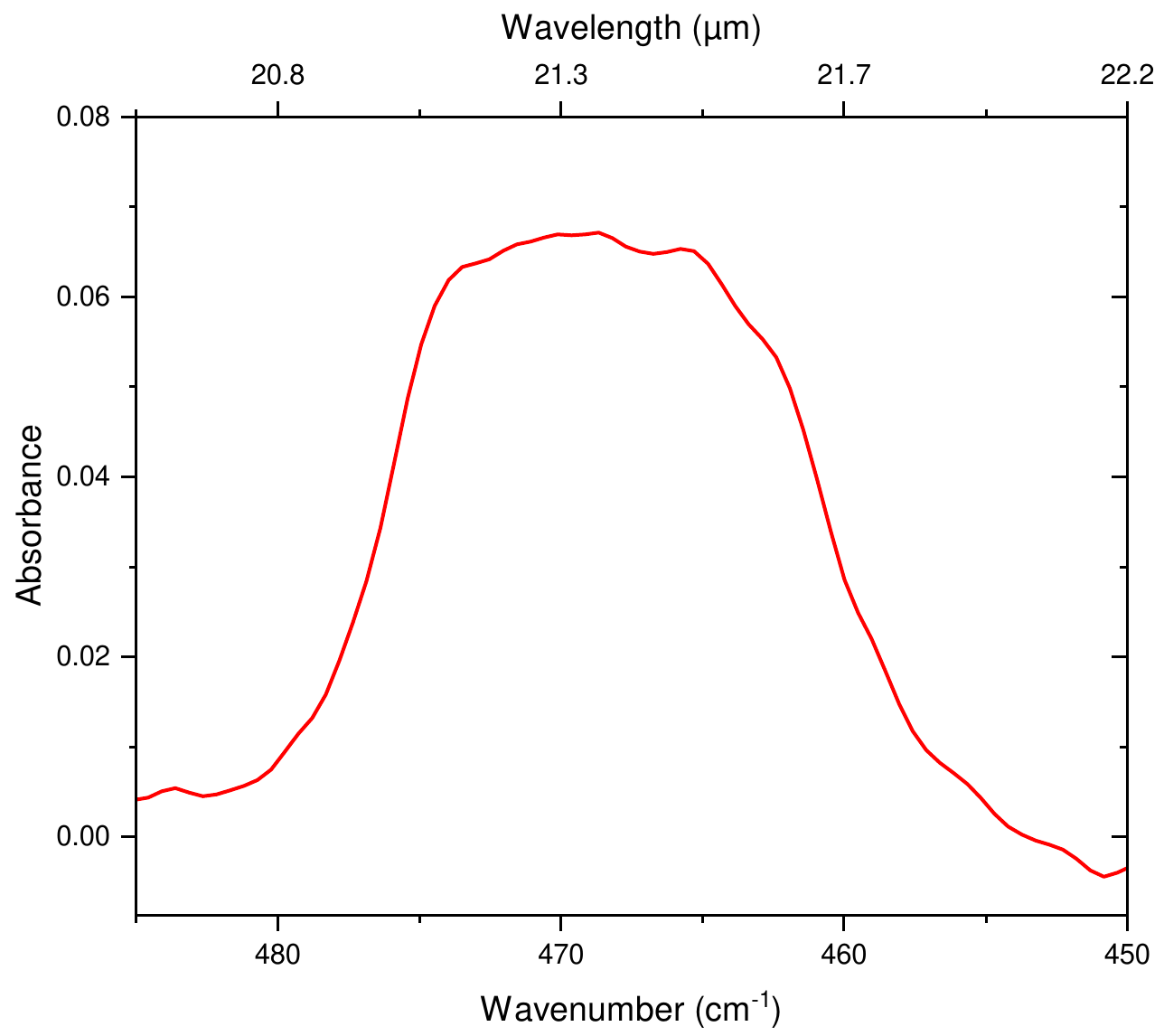}
    \caption{Most intense IR feature of S8 in the MIR corresponding to a solid S$_8$ pellet measured at the LSAIP from Centro de Astrobiología (CAB). The working resolution was 2 cm$^{-1}$. 
    }
    \label{fig:S8_lab}
\end{figure}

The estimation of the IR band strength of S$_8$ was carried out at the Centro de Astrobiología (CAB, CSIC-INTA, Spain). 
For this purpose, a S$_8$ pellet was formed by pressing 6$\pm$0.1 mg of S$_8$ into an area of $\sim$0.3 cm$^2$.
The IR spectrum of the resulting pellet was subsequently recorded in the sample compartment of a Bruker Vertex 70 Fourier-Transform IR (FTIR) spectrometer with a working resolution of 2 cm$^{-1}$. 
The most intense IR feature of S$_8$ in the MIR was detected at 21.36~$\mu$m (468~cm$^{-1}$), and is shown in Fig. \ref{fig:S8_lab}. 
The absorbance of the 468~cm$^{-1}$ IR band was numerically integrated using the \texttt{Origin} software (482~cm$^{-1}$ -- 451~cm$^{-1}$).
At the same time, the number of S$_8$ molecules per unit area (i.e. the S$_8$ column density $N$) was calculated using the S$_8$ molecular mass (256 amu) and Avogadro's number. 
Assuming a homogeneous distribution of the sulphur sample within the IR beam spot, a band strength of 1.5 $\times$ 10$^{-19}$ cm/molecule was obtained using Eq. \ref{eqn}. 

The main source of uncertainty in the estimation of the band strength was assumed to be the mass of the S$_8$ pellet, as some S$_8$ powder could have been lost during its preparation.
To ensure the reliability of the calculated band strength, two alternative methods were used to calculate the mass of the pellet. First, we measured its thickness and subsequently calculated its mass using the S$_8$ density (2.07~g~cm$^{-3}$), and assuming a surface of $\sim$0.3 cm$^2$. 
A second method made use of the observed fringes in the IR spectrum's baseline. The wavelength difference between two consecutive maxima was used to calculate the thickness of the pellet adopting a refractive index value of 1.94 for S$_8$
\citep{1985Icar...64..368S}. 
In both cases, the calculated mass of the pellet was close to the weight value of 6$\pm$0.1 mg .

\subsection{Laboratory measurements of CS$_2$ IR band strengths}\label{CS2}

In order to estimate the band strength of the main CS$_2$ IR feature at $\sim$6.65 $\mu$m ($\sim$1500 cm$^{-1}$), as well as a combination mode at $\sim$4.65 $\mu$m ($\sim$2145 cm$^{-1}$), at different temperatures, we evaluated the evolution of the IR spectrum of a pure CS$_2$ ice with temperature using the SPACE TIGER experimental setup at the Center for Astrophysics (CfA | Harvard \& Smithsonian, USA). 
SPACE TIGER consists of an ultra-high-vacuum (UHV) chamber with a base pressure of $\sim$2 $\times$ 10$^{-10}$ Torr at room temperature, similar to the pressure found in the interior of dense clouds. This setup is described in detail in \citet{2022ApJ...940..113M}. 
In particular, we calculated the relative integrated IR absorbance of the above-mentioned features at different temperatures with respect to their value at 80 K.  
The band strengths were subsequently estimated using the value at 77 K reported in \citet{1964JChPh..40..309Y} as reference, and assuming that the number of molecules per surface area did not change with temperature prior to the thermal desorption of the ice.

A pure CS$_2$ (anhydrous, Sigma-Aldrich, 99\%) ice sample was deposited on a CsI substrate at 8 K (achieved with a closed-cycle He cryostat, Model DE210B-g, Advance Research Systems, Inc.) and subsequently warmed-up by means of a 50 W silicon nitride cartridge heater rod (Bach Resistor Ceramics) until complete sublimation was attained at $\sim$130 K. 
The sample temperature was measured with a calibrated Si diode sensor with a 0.1 K relative uncertainty and controlled with a LakeShore 336 temperature controller. A heating rate of 2 K min$^{-1}$ was applied. 
The ice sample was monitored during warm-up with a Bruker 70v Fourier-transform infrared (FTIR) spectrometer in transmission mode, equipped with a liquid-nitrogen-cooled MCT detector. The spectra were averaged over 256 interferograms and collected with a resolution of 1 cm$^{-1}$ over the 5000$-$600 cm$^{-1}$ range. 

\begin{figure}[!htb]
    \centering
    \includegraphics[width=1.0\linewidth]{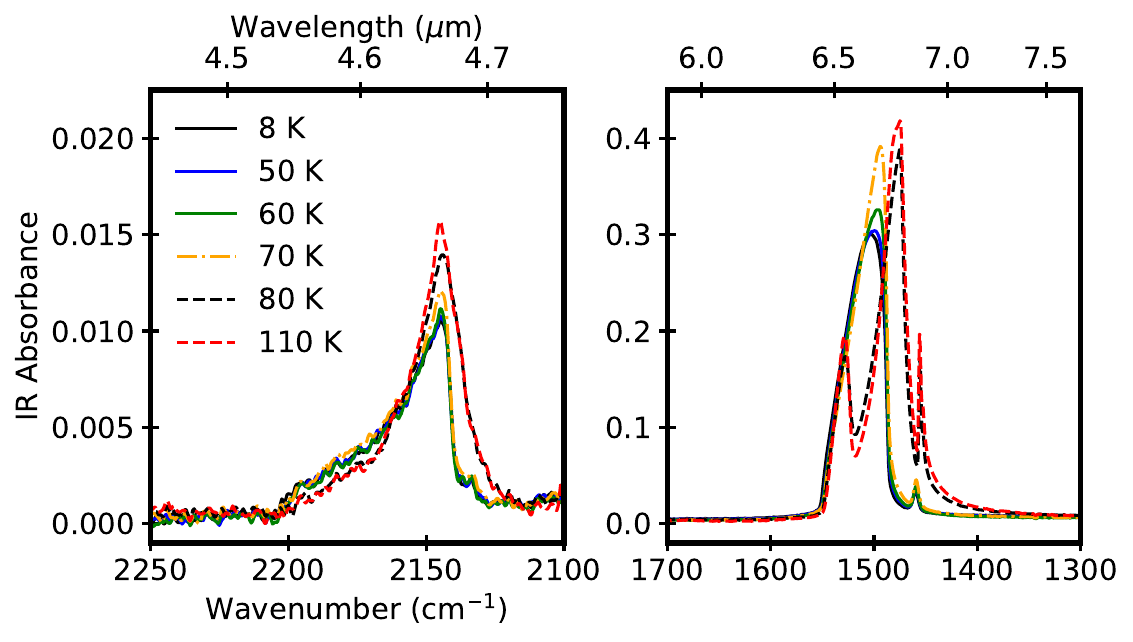}
    \caption{Temperature evolution of the $\sim$4.65 $\mu$m (left panel) and $\sim$6.65 $\mu$m (right panel) CS$_2$ IR features. The $\sim$6.85 $\mu$m feature corresponding to $^{13}$CS$_2$ was observed in the red wing of the main CS$_2$ IR band (right panel). Transition from amorphous (solid lines) to crystalline (dashed lines) structure was observed above 60 K.}
    \label{fig:IR_spec}
\end{figure}

\begin{figure}
    \centering
    \includegraphics[width=0.65\linewidth]{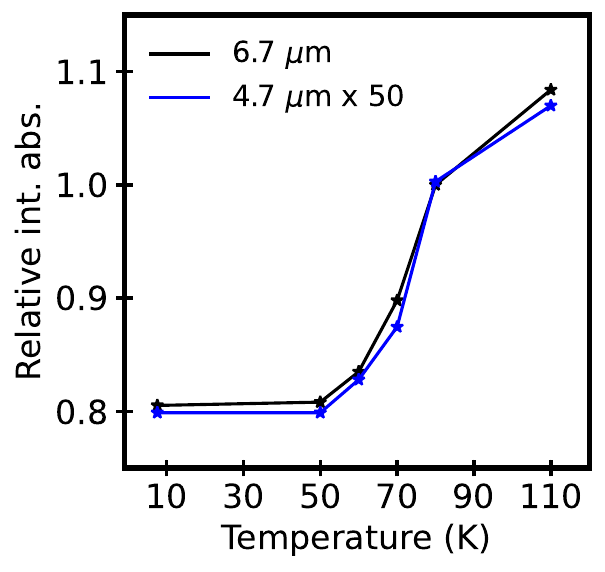}
    \caption{Relative integrated absorbance of the $\sim$4.65 $\mu$m (blue) and $\sim$6.65 $\mu$m (black) features with respect to the integrated absorbance of the main (6.65 $\mu$m) CS$_2$ IR band at 80 K as a function of the temperature.}
    \label{fig:IR_evol}
\end{figure}

Figure \ref{fig:IR_spec} shows the evolution of the $\sim$4.65 $\mu$m (left panel) and $\sim$6.65 $\mu$m (right panel) CS$_2$ IR features during warm-up of the ice sample. 
The $\sim$6.85 $\mu$m feature corresponding to $^{13}$CS$_2$ was observed in the red wing of the main CS$_2$ IR band.  
The 4.65 and 6.65 $\mu$m IR features were numerically integrated over the 2205$-$2110 cm$^{-1}$ and 1620$-$1270 cm$^{-1}$ ranges. 
The 6.85 $\mu$m feature was also numerically integrated over the 1460$-$1445 cm$^{-1}$ range, and subsequently subtracted from the integrated absorbance calculated for the 1620$-$1270 cm$^{-1}$ range in order to remove the contribution of the $^{13}$CS$_2$ molecules 
(note that the 6.85 $\mu$m integrated absorbance was $\le$3\% of the 6.65 $\mu$m feature). 
Figure \ref{fig:IR_evol} shows the relative integrated absorbance of the $\sim$4.65 $\mu$m (blue) and $\sim$6.65 $\mu$m (black) features with respect to the integrated absorbance of the main (6.65 $\mu$m) CS$_2$ IR band at 80 K as a function of the temperature. 
The spectra did not present significant differences in the 8$-$60 K range. Above 60 K the transition from amorphous to crystalline structure was evidenced by an increase in the integrated IR band absorbances. A split of the $\sim$6.65 $\mu$m feature in two peaks was observed once the transition was completed at $\sim$80 K (noting that the spectra were collected every 5 min during warm-up, so no IR spectrum was measured between 70 and 80 K). 
Desorption of the ice sample started above 110 K. 

As mentioned above, we assumed that the ice column density remained constant prior to the start of the thermal desorption at 110 K. Therefore, variations in the integrated absorbances of the IR features presented in Fig. \ref{fig:IR_evol} were due to changes in the IR band strengths of both features. 
Assuming the band strength of 1.3 $\times$ 10$^{-16}$ cm$^{-1}$ for the 6.65 $\mu$m CS$_2$ IR feature in a crystalline CS$_2$ ice sample at 77 K \citep{1964JChPh..40..309Y}, the band strengths for an amorphous ice at 8 K would be 1.1 $\times$ 10$^{-16}$ cm$^{-1}$/molecule and 2.1 $\times$ 10$^{-18}$ cm$^{-1}$/molecule for the $\sim$6.65 and $\sim$4.65 $\mu$m features, respectively. 

\subsection{Literature compendium of a fraction of S-bearing species present on the IR spectrum}

We summarise in Table~\ref{tab:spectral_data_S} the position and band strength associated with the main vibrational modes derived from laboratory experiments of S-bearing species found in the literature and from our experiments. 
These bands are mainly located in the wavelengths covered by MIRI (CS$_2$ at 6.66 $\mu$m, SO$_2$ at 7.49 $\mu$m, CS at 7.87 $\mu$m, SO at 8.85 $\mu$m, S$_3$ at 14.71 $\mu$m, S$_4$ at 15.04 $\mu$m and S$_8$ at 21.23 $\mu$m), with only three bands in the NIRSpec region (H$_2$S at 3.9 $\mu$m and OCS at 4.9 $\mu$m). 
The parameters presented in Table~\ref{tab:spectral_data_S} were extracted from pure ice experiments for OCS, H$_2$S, CS$_2$, and SO$_2$. 
The parameters for H$_2$S$_2$, SO, and CS were measured from mixtures with their S-bearing precursors (H$_2$S, CS$_2$ and SO$_2$). 
The S$_8$ band was measured at room temperature and H$_2$S$_2$ at 100 K. 

The S$_3$ and S$_4$ band strengths are unknown but we can make estimations based on their spectroscopic parameters.
O$_3$ presents an IR feature at 9.6 $\mu$m \citep{rao2012molecular}, which we used here as a first approximation. Considering that the reduced masses of the atoms are larger for S$_3$, the vibration frequency is lower (680 cm$^{-1}$ according to \citet{Brabson_1991}) for the S-S bond compared to 1040 cm$^{-1}$ for the asymmetric stretch of O$_3$. In addition, sulphur is not as electronegative as oxygen, meaning the vibration is less intense. In these conditions, the absorption feature of O$_3$ at 9.6 $\mu$m band strength would only serve to derive a overestimation to the column density of S$_3$. 
To estimate the value of the band strength, we have ran a quantum calculation to predict the relative band strengths in the gas phase (at the DFT/CAM-B3Lyp/6-311+G(d) level using the Gaussian 09 software suite \citep{Gaussian9}, assuming the harmonic approximation).
With this, we obtain both the relative line position $\nu$ and intensity I: for O$_3$, $\nu_{as, O_3}$ = 1284.92 cm$^{-1}$ and I$_{O_3}$ = 284.246 km/mole and for S$_3$, $\nu_{as, S_3}$ = 694.82 cm$^{-1}$, I$_{S_3}$ = 151.420 km/mole. The intensity ratio of the transitions of the two molecules is thus of $\sim$ 1.87. 
We can convert the intensity into band strength units by dividing by the Avogadro number. 
In \citet{1985Rao}, the authors derived the band strength of O$_3$ as a trace product of the irradiation of a CO$_2$:O$_2$ (1:1) mixture, with a value of about 1.4 $\times$ 10$^{-17}$ cm/molecule.
From the I$_{O_3}$ value computed with DFT, we find a value of 4.7 $\times$ 10$^{-17}$ cm/molecule which is about 3.3 times the value found in the literature. 
Using the conversion to band strength units, the S$_3$ value is about $\sim$ 2.5 $\times$ 10$^{-17}$ cm molecule$^{-1}$. 
Note that the band strength we derived from the quantum calculation is not the band strength measured over the integrated cross-section of the band and could involve large uncertainties. However we consider that this estimation is enough to test our model.
These results were independently confirmed with DFT simulations using the software CASTEP \citep{CASTEP} with a similar level of theory.

We applied the same method for S$_4$ by comparing it with O$_4$. In this case in particular, we have no information on the band strength of either of the features, but our previous computation with the band strength of the O$_3$ feature is close enough to apply it here.
Using the same quantum chemical calculation used for S$_3$, we first check the number of geometries accessible. For O$_4$, the lowest energy geometry is C$_{2h}$ and the next closest in energy, C$_{2v}$, is 1.5 eV higher. For S$_4$, our level of theory leads to two close-lying geometries: C$_{2v}$ at 0 eV and C$_{2h}$ at 0.02 eV \citep[also reported in][]{steudel2003}. C$_{2v}$ has no transitions in the MIRI wavelength range, while C$_{2h}$ has a strong transition at 16 $\mu$m. 
Here, we consider the C$_{2h}$ geometry as the two levels are extremely close energy-wise and within the detector range. Running the calculations, we get for O$_4$, at a position of $\nu_{as, O_4}$ = 1404.02 cm$^{-1}$, an intensity of I$_{O_4}$ = 899.634 km/mole and for S$_4$ at a position of $\nu_{as, S_4}$ = 619.43 cm$^{-1}$, an intensity I$_{S_4}$ = 300.625 km/mole. With these values, the ratio of intensities is about 3 and leads to band strengths of $\sim$ 1.5 $\times$ 10$^{-16}$ cm/molecule and $\sim$ 5.0 $\times$ 10$^{-17}$ cm/molecule for, respectfully, O$_4$ and S$_4$.

\begin{figure}
    \centering
    \includegraphics[width=\linewidth]{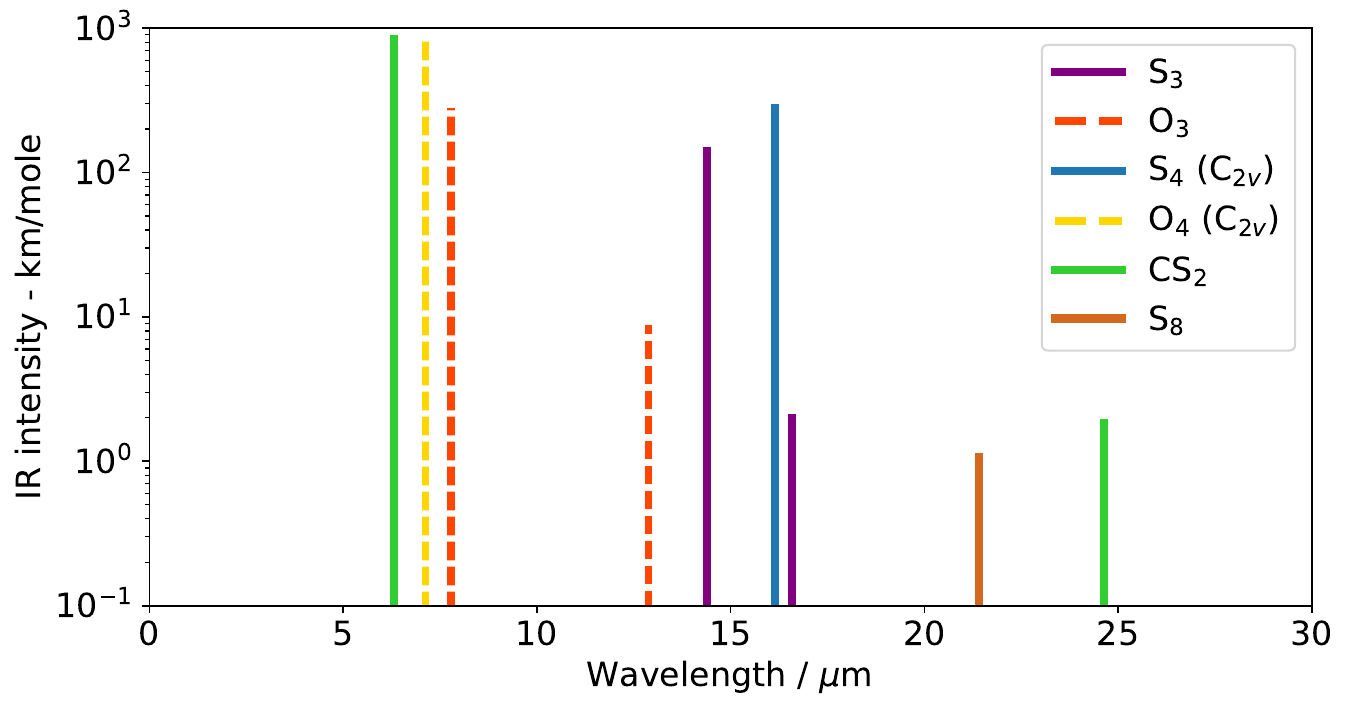}
    \caption{Calculated IR intensities of S$_3$, S$_4$, S$_8$, CS$_2$ compared to those derived for the equivalent oxygen-bearing allotropes O$_3$ and O$_4$. Harmonic approximation.}
    \label{fig_calcs}
\end{figure}

In Fig.~\ref{fig_calcs}, we plot the IR intensity as a function of the wavelength of the different vibrational modes of S$_3$, S$_4$, S$_8$ and CS$_2$ compared to O$_3$ and O$_4$. 
We note that a strong CS$_2$ vibration mode exists around 25 $\mu$m, not included in our detector coverage and, thus, not studied here. Similarly, S$_8$ is in the d4d point group and also presents transitions at $\nu$/I of 194/3.5, 249/6, 467/1, which are not covered in this study. 
One of the limitations to detect S-bearing species in ices in particular is due to the downshifted frequencies of the fundamental modes, adding another layer of difficulty to studying them in the lab.

Some other potential S-bearing species in the ices have been identified and their band strengths estimated in laboratory, but are not within the scope of this work. This is the case for H$_2$S$_2$, whose band strength was derived from the integrated area of its TPD peak ranging from 100 to 120 K \citep{2022AA...657A.100C} which is too uncertain. Ice bands of pure thiols have also been measured by \citet{Hudson_2018}, more precisely for CH$_3$SH, CH$_3$CH$_2$SH, CH$_3$CH$_2$CH$_2$SH, and (CH$_3$)$_2$CHSH, but these species are not expected to be meaningful S-carriers in our models. 

\section{The spectral position of sulphur bands in the IR spectrum}\label{SIS}

In this section, we present synthetic near- and mid- infrared (NIR and MIR) spectra, using these examples to discuss the difficulty of deconvolving the relatively weak spectral features of S-bearing species from the strong absorption features of other, often more abundant, molecules in the ice.
Section \ref{sec:synthicespec} presents the Synthetic Ice Spectrum Generator (SynthIceSpec) tool. Section \ref{simple_approach} lists the IR parameters extracted from laboratory experiments that were used in SynthIceSpec, along with a first evaluation of the potential overlap of IR bands of S-carriers with those corresponding to the major detected interstellar ice components.

\subsection{Synthetic ice spectrum generator}\label{sec:synthicespec}

To evaluate the detectability of S-bearing species in interstellar ices, we used a synthetic ice spectrum generator, SynthIceSpec\footnote{https://forge.oasu.u-bordeaux.fr/LAB/astrochem-tools/synthicespec.git}, to produce synthetic JWST spectra assuming reasonable input abundances. The code is described in detail in Taillard et al, (in prep).

In a nutshell, SynthIceSpec can produce ice absorption spectra, based on a first order approximation that vibrational modes of the molecular absorption in the solid phase can be represented as a sum of Gaussian functions. 
The band width, band strength and peak position of the different IR features are defined and derived based on spectra measured in laboratory experiments. 
For each simulated absorption feature, the optical depth of the band, $\tau(\nu)$, is obtained by multiplying the assumed column density of the associated molecule by the corresponding band strength distributed over a Gaussian profile with a given full width at half maximum (FWHM) over the wavenumber axis ($\nu$). 
The absorption spectrum is then obtained with $\exp{(-\tau(\nu))}$, after all the components are summed.
Three instruments (NIRCam, NIRSpec and MIRI on the JWST) have been parametrised so far, with their intrinsic characteristics (such as their resolution in different filters) pre-programmed. 
Column densities can be input directly by the user or be provided by output files from the Nautilus \citep{ruaud_gas_2016,Wakelam_2024} gas-grain model (or any other model with output files formatted in the same way), with column densities selected for any astrochemical time-step.
In the study presented here, the silicate signature and carbonaceous grains are not included and the absorption spectra are multiplied by a fiducial astronomical source spectrum represented here by a continuum flux of constant value (1, 0.1 and 0.01 mJy) across the wavelength range considered.

\subsection{Synthetic spectra model limitations}\label{model_limitation}

We have developed a methodology based on use of the SynthIceSpec synthetic tool to investigate the detectability of icy sulphur species in different environments using the JWST. However, its limitations must be taken into account when interpreting the synthetic spectra presented below.
To start with, despite its large database that is being updated with new laboratory data, some species (especially COMs in the fingerprint region) are still missing and perhaps some outdated values are used. 
Weaker features can often be observed in spectra of ices in SFR. In this case, we have not included the weak features $^{13}$CO \citep{Boogert_2002,Pontoppidan2003,2023NatAs...7..431M} and $^{13}$CO$_2$ \citep{Ehrenfreund_1997,Gerakines_1999,Pontoppidan_2008,Brunken_2024} nor the recently observed HDO \citep{Slavicinska_2024}. However, we do choose to include the recently observed dangling OH \citep{Noble2024}, which lie in the blue wing of the 3 $\mu$m water OH stretch. Isotopes bands parameters were not added in priority since they are not predicted in the base chemical network of Nautilus.

We emphasise here the significant role that mixing has in ices, affecting all bands present in a given ice sample. For the moment, although mixing with water at certain ratios for certain molecules is taken into account in the tool, the effect of mixing is not yet broadly implemented for all species in the synthesised spectra.
From a broader point of view, SynthIceSpec can accept chemical model outputs as inputs and, therefore, if we consider these models to be limited in their predictions, then it also affects the ice column densities we reproduce with the synthetic spectra.
From a physical point of view, it does not take into account any sort of radiative transfer or scattering, that would drastically change the baseline and push further away the identification of less abundant species. Ultimately, there is no consideration of the physical object itself, nor of the expected ice and grain modifications that give rise to profiles much more complex than a simple sum of Gaussian profiles (mixing, segregation, crystallisation, chemical complexification, grain size and shape effects on band profiles).
SynthIceSpec does not reproduce the variations that can be observed along the line of sight, nor give any constraint on which part of a protostar shell is being probed. This has major implications on the ice compositions, both in dense cold cores and in more advanced YSOs. All of these considerations mean that, independent of the synthetic spectra we present here, observations of very similar type of sources can vary. 
In summary, in this state, SynthIceSpec cannot be used to be compared with observations directly. However, even in its current state, it can put useful constraints on the observability of species with isolated absorption features, as we demonstrate below.
Future updates will introduce different parameters into SynthIceSpec (grain growth parametrisation, mixing effects, different ice component expected along the line of sight, etc.) to allow for a more direct comparison with observations. These parameters are described in an upcoming article dedicated to describing the tool (Taillard et al. in prep.).

\subsection{Ice mixing: major ice species are mixed, and sulphur-bearing species are mixed in water ice}\label{simple_approach}

\begin{figure}[ht]
    \centering
    \includegraphics[width=0.99\linewidth]{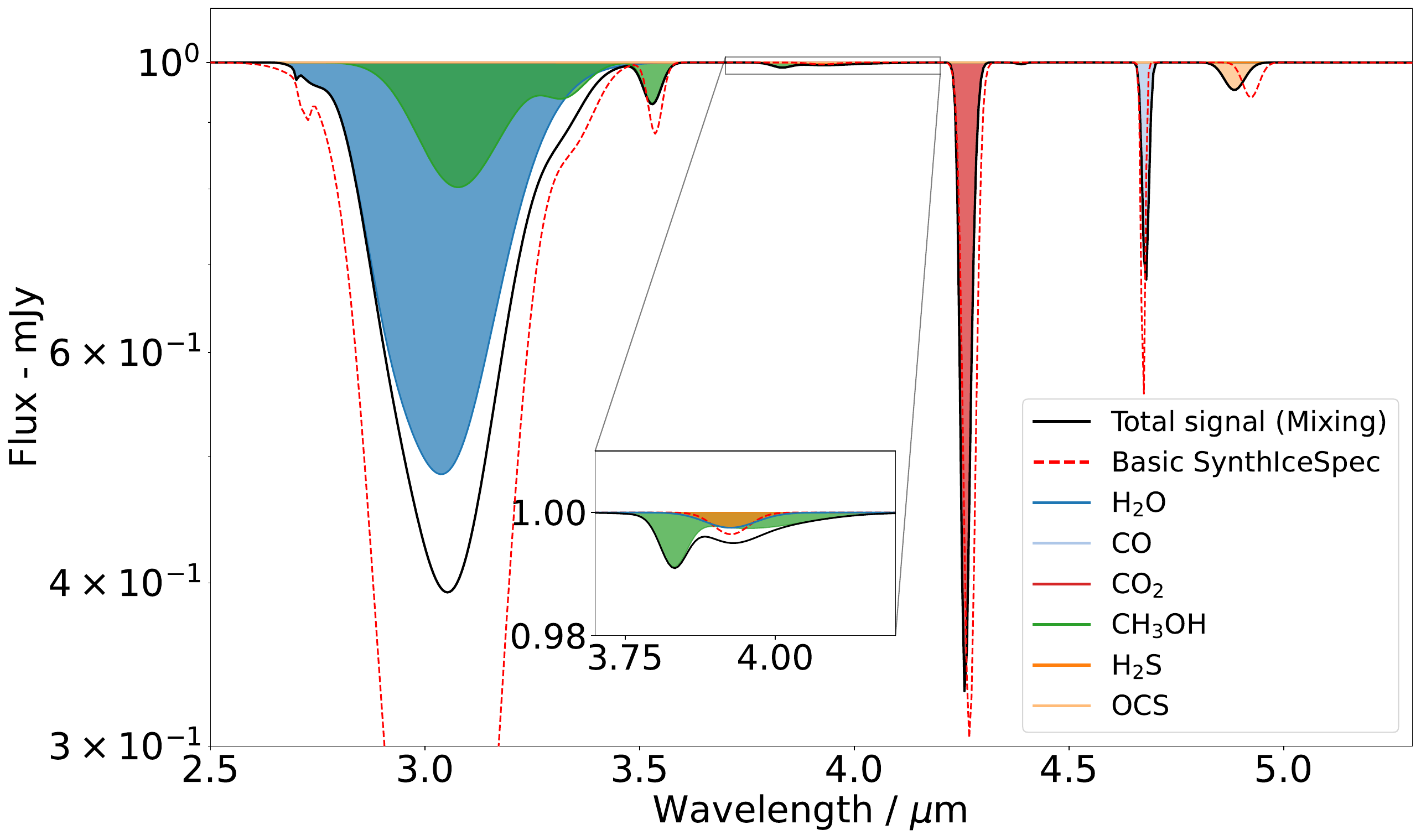}
    \caption{
Synthetic ice spectra of a `simple' ice composition, containing only the major ice species and S-bearing molecules in the NIRSpec wavelengths. 
The spectral information used for each species (band shapes, positions and band strengths) are listed in Table~\ref{tab:simple_approach_param}.
The column densities of H$_2$O, CO, CO$_2$, CH$_3$OH used to plot these spectra are, respectively, 3.0 $\times$ 10$^{18}$, 3.0 $\times$ 10$^{17}$, 3.0 $\times$ 10$^{17}$, and 5.0 $\times$ 10$^{17}$ molecules cm$^{-2}$, while we assume 1.0 $\times$ 10$^{16}$ molecules cm$^{-2}$ for all four S-bearing species (H$_2$S, OCS, SO$_2$, and CS$_2$). The zoom panel emphasises the low band strength of the band corresponding to H$_2$S, as well as its overlapping with the methanol combination mode. The red line is the signal that would be obtained with the most basic settings in SynthIceSpec (pure species with no mixing). In this wavelength range, of the S-bearing species, only OCS and H$_2$S present absorption features.}

    \label{fig:comp_H2OCOCH3OH_nirspec}
\end{figure}

\begin{figure}[h]
    \centering
    \includegraphics[width=0.99\linewidth]{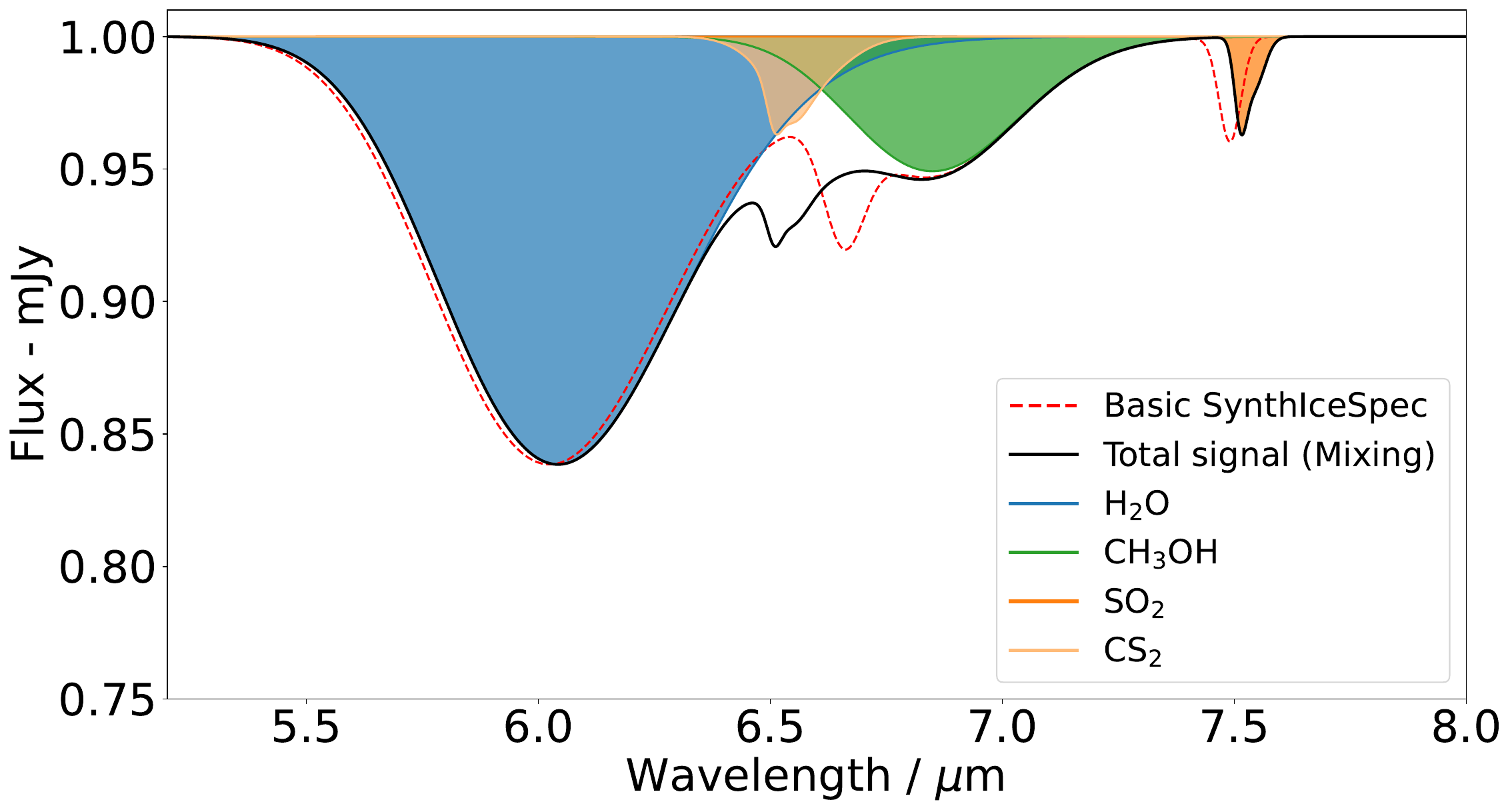}
    \caption{Same as Fig.~\ref{fig:comp_H2OCOCH3OH_nirspec}, but in the 5.2 to 8 $\mu$m region, as would be observed by MIRI. The zoom panel shows the weak SO$_2$ features. In this wavelength range, of the S-bearing species, only CS$_2$ and SO$_2$ present absorption features.} 
    \label{fig:comp_H2OCOCH3OH_miri}
\end{figure}

To test the detectability of sulphur-bearing species, we first compared the position, width, and intensities of their IR bands listed in Table \ref{tab:spectral_data_S} with those of the major ice components H$_2$O, CO, CO$_2$, and CH$_3$OH. This allows us to anticipate which bands of the S-bearing species would overlap with bands of major species present in most chemical environments in star-forming regions, before going on to look at a more complex ice composition. 
For this purpose, we used SynthIceSpec without adding any synthetic continuum emission, simply to check the possible overlap in the simplest case.
In Table~\ref{tab:simple_approach_param}, we list the different parameters derived to fit the Gaussian profiles for each band (bands position, intensity and width) with the associated ice compositions they were extracted from.
We extracted IR band parameters suitable for ice mixtures of H$_2$O, CO$_2$, and CH$_3$OH by manually fitting different Gaussian profiles to the 10 K spectrum of H$_2$O:CH$_3$OH:CO$_2$ (9:1:2) from \citet{1999A&A...350..240E}.
For the CO band, we used the ice mixture H$_2$O:CO (4:1) at 10 K from \citet{2007A&A...476..995B}. 
Additionally, rather than use the values for pure molecules presented in Table~\ref{tab:spectral_data_S}, we also took into consideration mixing effects of S-bearing molecules in water. The newly derived Gaussian parameters from mixed ices have been extracted from a H$_2$O:CS$_2$ (7:1) mixture at 5.5 K \citep{martin_domenech_2024_13134350}, H$_2$O:H$_2$S (7.5:1) mixture at 10 K \citep{jimenez-escobar_sulfur_2011} and of a H$_2$O:OCS (20:1) mixture at 10 K \citep{1993ApJS...86..713H} and are included in Table~\ref{tab:simple_approach_param}.
For SO$_2$ in water, we used the parameters measured by \citet{2003JMoSt.644..151S} for the band position, shape and width.
However, these authors did not measure the band strengths, therefore, for SO$_2$ in water, we used the value provided in \citet{2022ApJ...931L...4Y}, where it has been concluded that sulphur species' band strengths in amorphous water are not strongly affected at 10 K. 
For the H$_2$O:CS$_2$ (7:1), we also derived new water libration mode parameters, as these specific feature overlaps with the CS$_2$ one and is a bit shifted in position. 
The intensity was determined based on the assumption that the column densities derived for each spectrum (as provided in each original paper) is the full area under the curve of our Gaussian fit.
Figures \ref{fig:comp_H2OCOCH3OH_nirspec} and \ref{fig:comp_H2OCOCH3OH_miri} present spectra simulated from the simple ice composition for NIRSpec (parameters used are from the PRISM/CLEAR, covering 0.6 to 5.3 $\mu$m, R$\sim$100, the only available filter at the time in SynthIceSpec) and MIRI (MRS mode, all four channels, 4.9 to 27.9 $\mu$m, R$\sim$1460--3710) instruments, respectively. In the figures, we show only the wavelength ranges where the sulphur species are present: 2.5--5 $\mu$m for the NIRSpec part, where the strongest bands of H$_2$S and OCS are located and, similarly, 5--8 $\mu$m for the MIRI part, where the main absorption features of SO$_2$ and CS$_2$ are present.
Each species is plotted in a different colour, with the sum of all the components derived from the mixed ices (Table~\ref{tab:simple_approach_param}) plotted in black. In red, the same ice composition was used to compute a spectrum with the most basic version of SynthIceSpec (i.e. considering a simple sum of pure ice spectral features, without using the band parameters that include the effect of mixing). The species and their column densities used in the figures are as follows: H$_2$O (3.0 $\times$ 10$^{18}$ molecules cm$^{-2}$), CO (3.0 $\times$ 10$^{17}$ molecules cm$^{-2}$), CO$_2$ (5.0 $\times$ 10$^{17}$ molecules cm$^{-2}$), CH$_3$OH (5.0 $\times$ 10$^{17}$ cm$^{-2}$) corresponding to H$_2$O:CO$_2$:CO:CH$_3$OH 100:17:10:17 and the four S-bearing species (H$_2$S, OCS, SO$_2$, and CS$_2$) with the same column density (1.0 $\times$ 10$^{16}$ molecules cm$^{-2}$).

In this particular case, bands associated with OCS and SO$_2$ are both positioned in regions free of other spectral features while the bands of H$_2$S and CS$_2$ are overlapping some other bands. 
H$_2$S could also be considered to be an isolated band, unless methanol, in particular, is also present in the ice. 
The methanol combination mode at around 3.9 $\mu$m can appear in different mixtures and more specifically in a mixed ice presenting a high enough CH$_3$OH concentration \citep[see, in particular, Fig. 15 in][]{1999A&A...350..240E}, due to the hydrogen bonding interaction between the two molecules. 
When present, this particular vibrational mode can overlap with the H$_2$S band; when considering the latter's width and weak band strength, it can easily be blended with methanol.

The 6.66 $\mu$m band of CS$_2$ presents the strongest band strength of all the S-bearing molecules studied. However, the position of this absorption feature is not ideal, as it is in-between those of two very abundant molecules (H$_2$O and CH$_3$OH) and is deformed when in the presence of water. More mixtures would be needed to really quantify by how much it can be affected by the presence of other molecules.

Since the OCS band is quite isolated and its band strength high enough (about the same as for water ice, distributed over a much narrower bandwidth) to make its absorption band stand out from the baseline, it explains why its identification was made quite early. It can be noted that this molecule presents a strong dipole that does induce a red shift when mixed with water \citep{2022ApJ...931L...4Y}, where the peak position is moved by $\sim$ 15 cm$^{-1}$. This is an important parameter to take into account, considering that the band is very close to the detector's edge, which could easily challenge its proper detection.

Some species from Table~\ref{tab:spectral_data_S} have not been considered in our detectability cases. These molecules are: H$_2$S$_2$, SO, CS, S$_3$ and S$_4$.
CS$_2$, SO$_2$, OCS, and H$_2$S can be purchased in order to carry out experiments in the laboratory. Therefore, we can easily measure their IR spectra in realistic binary ice mixtures with H$_2$O and derive the required parameters, as shown in Table~\ref{tab:simple_approach_param}. SO, CS, and H$_2$S$_2$; however, these are not stable species on Earth, which prevens us from obtaining IR binary ice mixtures as were used in this study. For the sake of following the same methodology for all the considered species, we have not estimated the detectability of these three species.
In the cases of S$_3$ and S$_4$, despite their band strengths being estimated and their positions identified, we did not have an available spectrum for these two species to derive the Gaussian parameters we need. The spectra presented in \citet{Brabson_1991} show complex regions that need to be deconvolved, rather than derived by eye. Therefore, we decided to use the band width measured for S$_8$, although this introduces uncertainties on our results, presented later in this work for these two species.

\section{Detectability in three science cases}\label{detectability}

To follow the overall evolution and distribution of sulphur species during the star-formation process, we aim to evaluate the detectability of S-bearing molecules in three different types of source: a dense cloud, a low-mass young stellar object (LYSO) and a massive young stellar object (MYSO). 
For this purpose, we used the observed ice column densities compiled by \citet[][Table 2]{2015ARA&A..53..541B} for the major ice species in these environments.
We use the authors' mean value for each available molecular species as the column density input for generating our synthetic spectra, assuming they are case studies for the different steps of star formation that we cover. 
The molecules used in SynthIceSpec for these three cases are the following: H$_2$O, CO, CO$_2$, CH$_4$, NH$_3$, NH$_4^+$, OCN$^-$, CH$_3$OH, and HCOOH, with their respective column densities listed in Table~\ref{tab:col_dens}.
Since the publication of \citet{2015ARA&A..53..541B}, new column densities have been measured for certain molecules and it is important to take into account that the values used here could be updated (mostly for the dense cloud, where a few molecules are missing). Also, since these observations are from various sources with very different physical conditions, the mean value used might not reflect the diversity of encountered chemistry in similar environments. Therefore, the detectability threshold that we derive here could vary from the mean and would ultimately need to be readjusted from one specific source to another.

We have now identified which S-bearing molecules could be part of the depleted sulphur, based on theoretical studies, laboratory experiments, and observations. We can thus target the strongest IR features of these molecules with our synthetic ice spectra to see the feasibility of making such observations.
Using the band strength values listed in Table~\ref{tab:simple_approach_param}, we can compute the minimum column densities needed to consider a 5$\sigma$ detection limit within our synthetic ice spectra.
Thresholds were derived for each ice composition with the same approach: 
in particular, we added white noise with rms$_{mean}$ = 0.003 and 0.005 mJy to NIRSpec and MIRI spectra, respectively. This noise level is similar to that of the observations published by \citet{2023NatAs...7..431M}.
In parallel, we also produce a spectrum free of S-bearing species and noise (the `clear' spectrum). 
For each ice composition, we consider three different sources that have a constant flat continuum of 1, 0.1, and 0.04 mJy. This is a synthetic way to simulate different extinction levels that can be expected while observing star-forming regions. 
In total, we have nine different thresholds per species: three for each level of continuum for the three types of source object.
For each S-bearing species, we extracted the main feature, determined the ratio of the two spectra (the one containing S-bearing species with noise over the `clear' one), and iteratively increased the column density of the S-bearing species until reaching a signal-to-noise ratio of 5 on its absorption feature.

\begin{table*}[!htp] 
\centering
\caption{Column densities used in the synthetic spectra.}\label{tab:col_dens}
\begin{tabu}{|l|ccc|}
\hline
\multicolumn{1}{|c}{} & \multicolumn{3}{|c|}{Column density (molecules cm$^{-2}$)}                \\
\multicolumn{1}{|c}{} & \multicolumn{3}{|c|}{(\% H$_2$O)}                               \\
\hline
Molecule        & Dense cloud                      & LYSO                                      & MYSO     \\  
\hline
H$_2$O          & 9.3$\times$10$^{18}$           & 5.0$\times$10$^{18}$          & 5.0$\times$10$^{18}$           \\
CO              & 2.3$\times$10$^{18}$ (25.0\%)  & 1.3$\times$10$^{18}$ (25.0\%) & 3.5$\times$10$^{17}$ (7.0\%)  \\ 
CO$_2$          & 2.4$\times$10$^{18}$ (26.0\%)   & 1.3$\times$10$^{18}$ (26.0\%) & 9.5$\times$10$^{17}$ (19.0\%) \\  
CH$_3$OH        & 7.4$\times$10$^{17}$ (8.0\%)  & 3.0$\times$10$^{18}$ (6.0\%)  & 4.5$\times$10$^{17}$ (9.0\%)   \\ 
NH$_3$          & 0                              & 3.5$\times$10$^{17}$ (7.0\%)  & 3.5$\times$10$^{17}$ (7.0\%)  \\ 
NH$_4^+$        & 0                              & 4.0$\times$10$^{17}$ (8.0\%)  & 5.5$\times$10$^{17}$ (11.0\%) \\
CH$_4$          & 0                              & 1.5$\times$10$^{17}$ (3.0\%)  & 1.0$\times$10$^{17}$ (2.0\%)  \\ 
OCN$^-$         & 0                              & 2.0$\times$10$^{17}$ (4.0\%)  & 3.0$\times$10$^{16}$ (0.6\%)  \\
HCOOH           & 3.7$\times$10$^{17}$ (4\%)     & 1.0$\times$10$^{17}$ (2.0\%)  & 2.0$\times$10$^{17}$ (4.0\%)  \\  
CH$_3$CHO       & 0                              & 5.0$\times$10$^{17}$ (1.0\%)  & 2.7$\times$10$^{17}$ (5.5\%)  \\
H$_2$CO         & 0                              & 1.0$\times$10$^{17}$ (2.0\%)  & 2.5$\times$10$^{17}$ (5.0\%)  \\
\hline
\multicolumn{1}{|c}{} & \multicolumn{3}{|c|}{}                               \\
\multicolumn{1}{|c}{} & \multicolumn{3}{|c|}{Detection thresholds of sulphur species}                               \\
\multicolumn{1}{|c}{} & \multicolumn{3}{|c|}{(assuming rms level derived from \citet{2023NatAs...7..431M},}                               \\
\multicolumn{1}{|c}{} & \multicolumn{3}{|c|}{of 0.003 and 0.005 mJy for NIRSpec and MIRI respectively)}                               \\
\multicolumn{1}{|c}{} & \multicolumn{3}{|c|}{}  \\
\hline
H$_2$S (1 mJy)    &  4.6$\times$10$^{16}$ (0.5\%)  &  5.5$\times$10$^{16}$ (1.1\%)  &  5.0$\times$10$^{16}$ (1.0\%) \\
H$_2$S (0.10 mJy) &  3.5$\times$10$^{17}$ (3.8\%)  &  4.4$\times$10$^{17}$ (8.9\%)  &  3.9$\times$10$^{17}$ (7.9\%) \\
H$_2$S (0.04 mJy) &  6.4$\times$10$^{17}$ (6.9\%)  &  8.7$\times$10$^{17}$ (17.5\%) &  7.7$\times$10$^{17}$ (15.5\%) \\
OCS (1 mJy)       &  1.8$\times$10$^{15}$ (0.02\%) &  2.5$\times$10$^{15}$ (0.05\%) &  2.5$\times$10$^{15}$ (0.05\%)\\
OCS (0.1 mJy)     &  1.8$\times$10$^{16}$ (0.2\%)  &  2.2$\times$10$^{16}$ (0.45\%)  &  2.5$\times$10$^{16}$ (0.5\%)\\
OCS (0.04 mJy)    &  3.7$\times$10$^{16}$ (0.4\%)  &  5.0$\times$10$^{16}$ (1.0\%)  &  6.0$\times$10$^{16}$ (1.2\%)\\
CS$_2$ (1 mJy)    &  5.6$\times$10$^{15}$ (0.06\%) &  5.0$\times$10$^{15}$ (0.1\%)  &  4.0$\times$10$^{15}$ (0.08\%) \\
CS$_2$ (0.1 mJy)  &  3.7$\times$10$^{16}$ (0.4\%)  &  3.0$\times$10$^{16}$ (0.6\%)  &  2.5$\times$10$^{16}$ (0.5\%) \\
CS$_2$ (0.04 mJy) &  7.4$\times$10$^{17}$ (0.8\%)  &  5.0$\times$10$^{16}$ (1.0\%)  &  8.0$\times$10$^{16}$ (1.6\%) \\
SO$_2$ (1 mJy)    &  7.4$\times$10$^{15}$ (0.08\%) &  7.5$\times$10$^{15}$ (0.15\%) &  6.0$\times$10$^{15}$ (0.12\%)\\
SO$_2$ (0.1 mJy)  &  4.6$\times$10$^{16}$ (0.5\%)  &  6.0$\times$10$^{17}$ (1.2\%)  &  4.0$\times$10$^{17}$ (0.8\%)\\
SO$_2$ (0.04 mJy) &  9.3$\times$10$^{16}$ (1.0\%)  &  1.0$\times$10$^{17}$ (2.1\%)  &  8.5$\times$10$^{16}$ (1.7\%)\\
\multicolumn{1}{|c}{S$_8$}      &  \multicolumn{3}{|c|}{Undetectable assuming all cosmic S abundance locked in the ices}  \\
\hline\end{tabu}
\tablefoot{Column densities used in each synthetic spectrum, in parenthesis the abundance relative to water. Except for water, all values are extracted from Table 2 in \citet{2015ARA&A..53..541B}. The 5$\sigma$ detection thresholds are listed for each S-carrier at each continuum level for all three environments. These thresholds are computed from the rms value derived from \citet{2023NatAs...7..431M}.
}
\end{table*}

In all three star forming environments, we use the Gaussian parameters derived in Sect.~\ref{simple_approach} for H$_2$O, CO$_2$, and CH$_3$OH from \citet{1999A&A...350..240E} of the 10 K spectrum of H$_2$O:CH$_3$OH:CO$_2$ (9:1:2) mixture. The S-bearing molecules' band strengths, shapes, and positions are also those presented in our simple approach: their position and shape are from H$_2$O mixtures. 
For the other species considered (e.g. CH$_4$, NH$_3$, etc.), when available we used a mixing of the species with water, similarly to the S-bearing species. The parameters used are summarised in Table~\ref{tab:simple_approach_param}. 
The IR data used to fit the Gaussian profiles of the mixed ice absorption features were provided by different databases: Leiden Ice Database \citep{2022A&A...668A..63R}, Optical Constants Database (OCdb)\footnote{https://ocdb.smce.nasa.gov/}, and the Cosmic Ice Laboratory\footnote{https://science.gsfc.nasa.gov/691/cosmicice/}. The reference for each spectrum can be found in Table~\ref{tab:simple_approach_param}, alongside the ratio of the components in the ice mixture.
We did not include a 5$\sigma$ detection limit for S$_n$ species in this analysis, but these species are treated separately in Sect.~\ref{allotropes}.

We present in Figs.~\ref{fig:bg_star_no_continuum_sis}, \ref{fig:class0_no_continuum_sis}, and \ref{fig:classI_no_continuum_sis} the predicted synthetic ice spectra for, respectively, a dense cloud, a LYSO and a MYSO with the flux density (Jy) as a function of the wavelength ($\mu$m). For each figure, the black spectrum represents the synthetic observation without noise and without any S-bearing species (X$_{S-bearing}$ = 0 molecules cm$^{-2}$) and the red spectrum is the same composition but with 5$\sigma$ detection limits for each S-bearing species and synthetic instrumental noise (See Table~\ref{tab:col_dens} for the abundance relative to water). The figures combine both NIRSpec (from 2 to 5 $\mu$m; note that this range is not associated with any specific filter)
and MIRI ranges (5 -- 24.5 $\mu$m, again, these wavelengths do not correspond to the full bandpass, but are chosen to highlight ice absorption features).

\subsection{Dense cloud estimated detection thresholds}\label{models_predictions}

The column densities used for the spectra presented in Fig.~\ref{fig:bg_star_no_continuum_sis} are defined in Table~\ref{tab:col_dens}. In this example, the column density of water is set to 9.3 $\times$ 10$^{18}$ molecules cm$^{-2}$, as predicted by \citep{navarro-almaida_gas_2020} for TMC 1 (see also Appendix~\ref{appendixb}). This column density fits within the range of values for a dense cloud reported in \citet{2015ARA&A..53..541B}.

The H$_2$S feature at 3.92 $\mu$m has not yet been detected, but the authors of \citet{2023NatAs...7..431M} set an upper limit of 0.6\% H$_2$O in the Chamaeleon dense cloud from the observation of the highly extincted background star NIR38.
In our case, we also considered the detection feasible for abundances of 0.5\%, 3.8\% and 9.3\% H$_2$O, respectively, for a constant flux of 1 mJy, 0.1 mJy and 0.04 mJy.

For the OCS feature at 4.9 $\mu$m, the detection could be challenged by the type of star observed, by the presence of photospheric features, especially at low column densities. Since we do not account for these gas-phase line emissions, our estimated detection thresholds are quite low: 0.02\%, 0.2\% and 0.4\% relative to water, respectively, for a constant flux of 1 mJy, 0.1 mJy and 0.04 mJy. 

The CS$_2$ feature at 6.7 $\mu$m, with our updated band strength, would seem detectable with an abundance relative to water of 0.06\%, 0.4\% and 0.8\% with, respectively, a constant flux of 1 mJy, 0.1 mJy and 0.04 mJy.

The SO$_2$ band at 7.5 $\mu$m, similarly to H$_2$S, has not been solidly detected. \citet{2023NatAs...7..431M} reported upper-limits of 0.025-0.04\% with respect to water. Our threshold is estimated as 0.08\%, 0.5\% and 1.0\% relative to water, respectively, for a constant flux of 1 mJy, 0.1 mJy and 0.04 mJy. We emphasise here that, despite the 'distinct' appearance of the feature in teal in Fig.~\ref{fig:bg_star_no_continuum_sis}, the band could be superimposed on those of other species, which are in our simulation set to a column density of 0.

\subsection{LYSO estimated detection thresholds}

LYSOs could present interesting ice features, since their gas-phase sulphur component has been studied thoroughly, with SO$_2$ being the most abundant molecular sulphur species \citep{Taquet_2020,Codella_2021}.

To our knowledge, there have not been any published JWST observations with NIRSpec of LYSO focussed on the inventory of ice species; therefore, we cannot compare our estimated detection thresholds with observations. \textit{Spitzer} observations have not shown any evidence of the presence of the considered sulphur species \citep{2008ApJ...678..985B} on the 5-8 $\mu$m range and no upper limits on H$_2$S have been estimated with ground-based telescope in LYSOs.
Our synthetic spectra are presented in Fig.~\ref{fig:class0_no_continuum_sis}.
The water column density used for this source was set arbitrarily to 5.0 $\times$ 10$^{18}$ molecules cm$^{-2}$. For H$_2$S, we predicted a possible detection at 1.1\%, 8.9\% and 17.5\% relative to water, respectively, for a constant flux of 1 mJy, 0.1 mJy, and 0.04 mJy.

With our ice spectra simulations, we estimated OCS detection thresholds of 0.05\%, 0.45\%, and 1.0\% relative to water, respectively, for a constant flux of 1 mJy, 0.1 mJy, and 0.04 mJy.
We determined possible thresholds for CS$_2$ of 0.1\%, 0.6\% and 1.0\% relative to water, respectively, for a constant flux of 1 mJy, 0.1 mJy, and 0.04 mJy.
Our SO$_2$ estimated thresholds are 0.15\%, 1.2\%, and 2.1\% relative to water, respectively, for a constant flux of 1 mJy, 0.1 mJy, and 0.04 mJy. \

\subsection{MYSO estimated detection thresholds}

For the MYSO simulated spectra, the water column density is also arbitrarily set at 5 $\times$ 10$^{18}$ molecules cm$^{-2}$. This value is within the range of observations reported in \citet{1995ApJ...449..674P} (0.5-50 $\times$ 10$^{18}$ molecules cm$^{-2}$) and \citet{2022ApJ...941...32B} (0.76-11.5 $\times$ 10$^{18}$ molecules cm$^{-2}$).

For H$_2$S, our thresholds are estimated at 1.0\%, 72.9\%, and 15.5\% relative to water, respectively, for a constant flux of 1 mJy, 0.1 mJy and 0.04 mJy. For OCS, we estimated thresholds of 0.05\%, 0.5\%, and 1.2\% relative to water, respectively, for a constant flux of 1 mJy, 0.1 mJy, and 0.04 mJy.

Our CS$_2$ detection thresholds in a MYSO are expected to be at 0.08\%, 0.5\% and 1.6\% relative to water, respectively, for a constant flux of 1 mJy, 0.1 mJy, and 0.04 mJy.
We estimated the detection thresholds for SO$_2$ of 0.12\%, 0.8\%, and 1.7\% relative to water, respectively, for a constant flux of 1 mJy, 0.1 mJy, and 0.04 mJy.

\section{Astrochemical implications}\label{discussion}

This section discusses the physical conditions and the ice chemistry expected in the different interstellar regions, and identifies the circumstances (ice composition) that would favour the formation of relatively large and detectable quantities of S-bearing ices in the three different environments.

\subsection{Favourable dense cloud environment to detect S-bearing molecules}\label{sub_dense_core}

Dense cloud environments can vary a lot from a cloud to another, with ice composition variation within the same cloud \citep[as was observed in][]{2017MNRAS.467.4753N}. 
Ice composition is highly influenced by the dynamics of the core and, more precisely, by the so-called `catastrophic CO freeze-out' \citep{bacmann_degree_2002}, where CO in the gas-phase is heavily depleted onto the grains. The depletion will cause a shift from a highly H$_2$O-rich surface to a CO-rich layer (leading to additional CO$_2$ formation). 
The chemistry is therefore critically impacted and could make the different detections challenging, depending on the regime the ices are experiencing, as new molecules will form.

In the case of the H$_2$S feature at 3.92 $\mu$m, we determined detection thresholds of 0.5\%, 3.8\%, and 6.9\% (w.r.t. to the H$_2$O abundance, respectively) for a constant flux of 1 mJy, 0.1 mJy, and 0.04 mJy. The detection thresholds we derived fall within the cometary abundances relative to water and within the gas-grain model predictions made by \citet{navarro-almaida_gas_2020}. 
With the predicted methanol column density and mixing used to derive the shape of the combination mode, H$_2$S would be overlapping, but strong enough to be visible. 
However, this is considering that the water band at 3.0 $\mu$m does not present a strong red-wing due to light scattering as seen, for example, in the recently published Ice Age spectra \citep{2023NatAs...7..431M,Dartois2024}.
Scattering effects have been observed ever since the first H$_2$O ice observations were made \citep{Gillett_Forrest_1973}. It is thus highly expected that in a dense cloud environment, the NIR observations are likely to be affected, with the H$_2$S feature blended in the red wing of water \citep{Potapov2021}.

H$_2$S is a central molecule in the chemistry of sulphur, with its high reactivity allowing it to efficiently reform SH through H abstraction.
\citet{2022A&A...659A.100E} experimentally studied the hydrogenation of H$_2$S molecules in the presence of CO and O$_2$ in prestellar environment conditions (T $<$ 30 K).
H$_2$S molecules readily react under these conditions, leading to the formation of organo-sulphur species already at low temperatures (such as H$_2$CS, CS, OCS, CCS, CH$_3$SH...). In any case, a fraction of the H$_2$S could remain unreacted and constitute a column density high enough to be detected. In particular, the authors raised two possible scenarios that could explain how 
H$_2$S could still be present in the ices at detectable abundances. The first is that the H$_2$S/CO ratio is too high and upon reaching a saturation limit, reactants would then be limited by the amount of available H atoms remaining on the surface. 
The second scenario would be that the amount of O$_2$ within the ices can negatively impact the reactivity of H$_2$S, competing with CO and H (where these two species would react more easily with O$_2$).
Interestingly, the first scenario could occur in TMC-1, where the measured C/O ratio is high in the gas-phase and with a strong sulphur depletion factor \citep{2013ChRv..113.8710A}. It could constitute the optimal conditions for more abundant H$_2$S ice and possibly lead to a very important sulphur-driven chemistry. However, the non-detectability of O$_2$ (no infrared transitions) hampers the resolution of this question, as it is difficult to obtain the oxygen budget within the ices. 
Laboratory experiments have shown that exposed to H atoms (Santos et al. 2024), the radical itself can react with carbon chains, such as C$_2$H$_2$ and form more complex thiols (ie, molecules with a -SH functional group). SH is also a central component in the formation of OCS, as it readily reacts with CO and could be strongly favoured during the catastrophic CO freeze-out (Santos et al. 2024). 
Other laboratory experiments proved that when exposed to radiation similar to that present in clouds, H$_2$S can efficiently lead to the formation of OCS, SO$_2$ and CS$_2$, depending on its chemical environment \citep[CO and/or CO$_2$ rich ices,][]{2008ApJ...684.1210F,2010A&A...509A..67G} but also could be the starting point of S-chains \citep{jimenez-escobar_sulfur_2011,2015ApJ...798...80C}. Spontaneous acid-base reactions involving S-bearing molecules have been recently highlighted in laboratory experiments \citep{Slavicinska_2024,Vitorino_2024}, with the salt NH$_4$SH formed through the reaction between NH$_3$ and H$_2$S that can be formed at low temperature ($\sim$ 10 K). The salt could play an important role in the non-detection of H$_2$S, as an important part of it could be locked in this form. Both experiments and models have also shown the importance of accounting for H$_2$S destruction channels to mitigate for the overproduction of H$_2$S in models of clouds \citep{Oba_2018,2019A&A...624A.108L}. Taking into account its fast reactivity, H$_2$S is playing an important role in the formation of other S-bearing species, with perhaps a smaller amount remaining pure in the ices and not being enough to allow for a proper detection.

For OCS, we determined thresholds for the feature at 4.9 $\mu$m of 0.02 \%, 0.2\%, and 0.4\% relative to water, for a constant flux of 1 mJy, 0.1 mJy and 0.04 mJy, respectively. 
The thresholds at 1 mJy are within the range detected in comets and both 1 and 0.1 mJy are within the values predicted by the models of \citet{navarro-almaida_gas_2020}. The threshold derived for the faintest continuum of 0.04 mJy is within a factor 6 (4) higher than the cometary observations (model predictions).
The composition of cometary comae may not be representative of the interstellar ice composition in all kinds of environments. Indeed, the OCS abundances can be higher in cold environments were CO is frozen onto the grain surfaces, promoting the formation of molecules such as CO$_2$, CH$_3$OH, and OCS \citep{2015ARA&A..53..541B}. Higher OCS abundances can enable its detection, even in 
regions with a low continuum flux.

For the CS$_2$ feature at 6.7 $\mu$m, our thresholds are of 0.06\%, 0.4\%, and 0.8\% relative to water, for a constant flux of 1 mJy, 0.1 mJy, and 0.04 mJy, respectively. The lowest threshold is almost 10 times higher than the highest observed values in comet 67P \citep{calmonte_sulphur-bearing_2016}.
At low temperatures, CS$_2$ can be formed in ices from CS + HS or from HCS + S. It has been shown experimentally that CS$_2$ can also be formed from cosmic-ray interactions with an H$_2$S mantle \citep{2010A&A...509A..67G} 
and secondary UV processing of ices containing H$_2$S \citep{2015ApJ...798...80C}. As the formation path in models is not yet well defined for CS$_2$, it is not formed efficiently. 
\citet{2019A&A...624A.108L} added a few reaction paths for the formation of CS$_2$ to their model and it seems to be formed at high density ($>$ 10$^5$ cm$^{-3}$) after long integration times (10$^{5}$ yr), but representing only a few percent of the total sulphur found in the ices. Considering the slow chemistry of CS$_2$, the formation of a reservoir abundant enough to be detected seems unlikely, unless some reaction pathways remain hidden or underestimated.

For the SO$_2$ band at 7.5 $\mu$m, we estimated detection thresholds at 0.08\%, 0.5\% and 1.0\% relative to water, for a constant flux of 1 mJy, 0.1 mJy, and 0.04 mJy, respectively. The formation of SO$_2$ is thought to be enhanced by the depletion of sulphur from the gas phase \citep{2019A&A...624A.108L}, which, in the case of a cloud similar to TMC-1, could have already happened during the translucent phase. However, due to the high binding energy of O \citep{Ward2012,Minissale2016}, SO$_2$ formation by SO + O is likely to be slow, meaning it could be formed at earlier stages of the molecular cloud (e.g. translucent phase), when the grain is warm enough to allow for the diffusion of oxygen (T$\sim$ 12 K) or mainly formed in the gas-phase and then depleted onto the grain (or from a direct Eley-Rideal mechanism, involving its precursor SO). 
Considering the limited chemistry as well as the perturbation of this band due to interaction with the features of other ice species, such as COMs and the H$_2$O bending mode, SO$_2$ detection seems to be one of the most challenging of the S-bearing species we consider in this particular environment. 
An absorption feature of OCN$^-$ at 7.5 $\mu$m could be expected to overlap, but according to laboratory experiments \citet{Hudson_2001,vanBroekhuizen_2004,Bennett_2010} and chemical modelling \citep{Theule_2011}, this molecule is efficiently formed through UV-irradiation of HNCO-rich ices, which are not expected to be abundant in dense cloud conditions. We do not expect a strong overlap between SO$_2$ and OCN$^-$ features in this case.
Another challenge arises from the difficulty of defining the baseline in general, that could highly affect the column density derived or simply the identification of a band.

\subsection{Favourable LYSO environment to detect S-bearing molecules}

As the core undergoes collapse and heats up, a fraction of the species locked in the ices will be sublimated and released into the gas-phase, unlocking a subsequent and more temperature-efficient chemistry. The rising grain temperature will lead to different chemical reactions on the surface.

The upper limit thresholds we derive for H$_2$S in a LYSO are 1.1\%, 8.9\% and 17.5\% relative to water, respectively, for a constant flux of 1 mJy, 0.1 mJy and 0.04 mJy. The H$_2$S reservoir, like other species presented here, would highly depend on the prestellar stage, where, depending on the scenario presented previously for H$_2$S in Sect.~\ref{sub_dense_core}, could strongly impact its abundance in warmer environments \citep{navarro-almaida_gas_2020}. 
It is also worth noting that H$_2$S is volatile and can easily sublimate at T $>$ 100 K \citep{hudson2018}. If mixed with water, a fraction will first desorb near 82 K in the lab and the rest codesorbs with water at up to 164 K \citep[see for instance Fig. 4 in][]{jimenez-escobar_sulfur_2011}. 
In \citet{martindomenech2014}, the authors computed a conversion factor from lab temperatures to circumstellar temperatures, for which a grain at a certain distance would be heated by the central object, corresponding to a factor of 0.62. Applying this factor would translate to the first desorption happening at a circumstellar temperature of T $\sim$ 50 K and the rest would codesorb around T $\sim$ 100 K.
This implies that its detection towards hot cores starts to be challenged by whether or not it can still be present in the solid phase.

For OCS, we determined thresholds of 0.05\%, 0.45\% and 1.0\% relative to water, respectively, for a constant flux of 1 mJy, 0.1 mJy and 0.04 mJy. OCS is expected to form from many other sulphur species, as mentioned before, and can also be a product of reprocessing by radiation. In \citet{vidal_reservoir_2017}, the authors were able to reproduce OCS observations of the deeply embedded protostar W33A, using an updated sulphur chemistry network. This was done considering an elemental sulphur abundance close to the cosmic one, which could be interpreted as an environment where almost all the sulphur would have depleted onto the grain and be locked in the ices. \cite{1997ApJ...479..839P} have shown with their observations that OCS can be present in abundances up to 5\% relative to CO ice, suggesting that the S-atom addition to CO is an efficient formation route. More energetic processes could also play an important role in the production of the molecule. \citet{2008ApJ...684.1210F} and \citet{2010A&A...509A..67G} showed that heating of CO:H$_2$S = 5:1 and CO:SO$_2$ ices can lead to the formation of the molecule. Probing the irradiation history of multiple sources could help to better constrain the formation route.

In the case of CS$_2$, we determined detection thresholds of 0.1\%, 0.6\% and 1.0\% relative to water, respectively, for a constant flux of 1 mJy, 0.1 mJy and 0.04 mJy. One way of forming CS$_2$ that has been explored in the lab is through irradiation of H$_2$S in CO ices \citep{2008ApJ...684.1210F}, that could occur in a class 0 environment, as the proto-star will irradiate the shell surrounding it. Similarly to the estimated detection threshold in dense clouds, CS$_2$ formation is not well constrained and would also need updated data in chemical models for protostars. It is difficult to rely on the value we estimated as we cannot even compare it with models.

The thresholds for SO$_2$ we have determined are 0.15\%, 1.2\%, and 2.1\% relative to water, respectively, for a constant flux of 1 mJy, 0.1 mJy, and 0.04 mJy. Similarly to CS$_2$, SO$_2$ can be formed through irradiation of H$_2$S, but in a CO$_2$ ice (as studied in experiments), due the fragmentation of the carbon dioxide \citep{2008ApJ...684.1210F}.
In this particular stage, where the grain is not warm enough to sublimate all the bulk ice, the SO$_2$ abundance in ices could be amplified as the grain would undergo irradiation processes. A prerequisite would be that the prestellar core was able to form a particularly rich CO$_2$ mantle. CO$_2$ ices undergoing segregation have been previously observed \citep[with \textit{Spitzer},][]{Pontoppidan2008,Zasowski2009} in YSO, supporting this scenario. 
In this particular case, the abundance of OCN$^-$ is expected to be higher than in the dense cloud environment, which means the absorption feature of OCN$^-$ at 7.5 $\mu$m would be affecting the identification and determination of SO$_2$ column density.

\subsection{Favourable MYSO environment to detect S-bearing molecules}

Just as for the LYSO environment, the ice composition around MYSOs is altered by the heating of the surrounding material, with expected ice thermal processing. 
Most volatile species would desorb directly from the ice surface while some, due to their higher sublimation temperature, would remain locked in the ice. For example, it is expected that CO would be evaporated by protostellar heating whereas CO$_2$ would be distilled in the ices \citep{2011ApJ...740..109O}. It was previously observed in MYSO with \textit{ISO} \citep{Ehrenfreund1998} and \textit{Spitzer} \citep{Seale2011}.
The IR features would probably be strongly affected in this environment and SynthIceSpec is probably the least accurate here, as thermal processing is not implemented in any sort of form. If we assume that the observation is done towards the cold envelope, then thermal processing would not affect the spectral features too much.

In the MYSO environment, we determined detection thresholds for H$_2$S of 1.0\%, 7.9\% and 15.5\% relative to water, respectively, for a constant flux of 1 mJy, 0.1 mJy and 0.04 mJy. Except for the brightest continuum value, that is fairly reasonable, the two other thresholds are high in comparison to \citet{1991MNRAS.249..172S}, who placed upper-limits on H$_2$S of $<$ 0.3-1\% relative to H$_2$O (in three high-mass protostars). 
Other upper-limits for H$_2$S in 2 protostars were also provided in \citet{jimenez-escobar_sulfur_2011}. In this study, the authors estimated these values to range from 0.13 to 0.7\% H$_2$O, using an UV irradiated H$_2$S:H$_2$O (13:100) laboratory ice mixture as a template. These upper limits fit within our simulated dense cloud and LYSO thresholds, but not for our MYSO environment.
Considering that over 30 K, H atoms are quickly desorbed from ice mantles and that even H + H does not efficiently react \citep{2007JChPh.127n4709A}, the reaction SH + H $\longrightarrow$ H$_2$S would then be fairly less effective in the MYSO temperature range. 
\citet{jimenez-escobar_sulfur_2011} showed that pure H$_2$S desorbs at a temperature of 82 K, meaning it would only be present at the edge of the class I shell. In their experiments, H$_2$S in a H$_2$O ice codesorbs over a temperature range of 130--170 K. By applying the factor derived from \citet{martindomenech2014} mentioned earlier, the first H$_2$S desorption event would occur at $\sim$ 50 K, and the expected codesorption with water would be at $\sim$ 80 -- 105 K in the MYSO environment.
It is expected that H$_2$S will be photolysed and lead to the formation of several species (including H$_2$S$_2$, HS, S$_2$...) and more oxygenated species if it is in a H$_2$O mantle (SO$_2$, HSO$_4^-$...) \citep{2022A&A...659A.100E,jimenez-escobar_sulfur_2011}.

The OCS thresholds are of 0.05\%, 0.5\% and 1.6\% relative to water, respectively, for a constant flux of 1 mJy, 0.1 mJy and 0.04 mJy. Like in the LYSO environment, radiation processing would occur in a MYSO environment and, similar to the lower-mass star forming region, both the presence of H$_2$S or SO$_2$ in CO ices and the heating of the mantle would result in the formation of OCS. Our thresholds at 1 and 0.1 mJy are within the values estimated in the MYSO survey from \citet{2022ApJ...941...32B}, ranging from 0.04 to 0.78\% with respect to water.

For CS$_2$, the thresholds are 0.08\%, 0.5\%, and 1.6\% relative to water, respectively, for a constant flux of 1 mJy, 0.1 mJy, and 0.04 mJy. From chemical models, CS$_2$ does not seem to have any chemical pathways that could be favoured in comparison to other molecules. The irradiation, as presented in the LYSO case, could possibly form CS$_2$, but it is still difficult to determine to what extent it would be favoured in a MYSO environment.
Consequently, it is difficult to assess CS$_2$ as a candidate for detection, Although the CS$_2$ thresholds seem to be reasonable compared to other ice species, they do not match the abundances observed in comets \citep[0.003–0.024\% w.r.t. water][]{LeRoy2015}.

Finally, for SO$_2$ in the MYSO case, our thresholds are 0.12\%, 0.8\% and 1.7\% relative to water, respectively, for a constant flux of 1 mJy, 0.1 mJy, and 0.04 mJy. The low flux threshold is within the range of values ($<$ 0.9 - 1.4\% w.r.t. to water) derived from the MYSO survey conducted by \citet{2004ApJS..151...35G}.
Our estimated values are a bit higher compared to the column densities derived ($\sim$ 0.1 - 0.2\% w.r.t. to water) from the possibly identified band in \citet{Rocha2024}. The authors show a very complicated spectral region surrounding the SO$_2$ band where they used local fits to measure its column density. 
As \citet{2022ApJ...941...32B} showed in a large fraction of their sample, MYSOs can present an excess of OCN$^-$ in the ice, whose absorption feature at 7.62~$\mu$m could overlap with that of SO$_2$ and hinder its detection. 
We note that in both \citet{Boogert_1997} and \citet{Rocha2024}, the authors used the mixture CH$_3$OH:SO$_2$ (11:1) measured in the first study, which causes a slight red-shifting of the SO$_2$ feature (similar to the H$_2$O:SO$_2$ mixing we are using) compared to its pure spectrum. 
Based on experimental results \citep{Martin-Domenech2024}, at least a fraction of SO$_2$ could be formed in photo-processed CO ice. 
As such, our team has recently measured the IR spectrum of a CO:SO$_2$ (16:1) ice mixture that exhibits a blue-shift as compared to both pure SO$_2$ and to the two ice mixtures described above \citep[][R. Martín-Doménech in prep]{2010A&A...509A..67G}. This has important implications: whether the SO$_2$ ice is locked in methanol and/or water matrices or in CO matrices, its identification and detection would be impacted, it will be further investigated in a follow-up study.

\subsection{sulphur allotropes and the (semi-)refractory S-bearing species}\label{allotropes}

\begin{figure}
    \centering
    \includegraphics[width=0.99\linewidth]{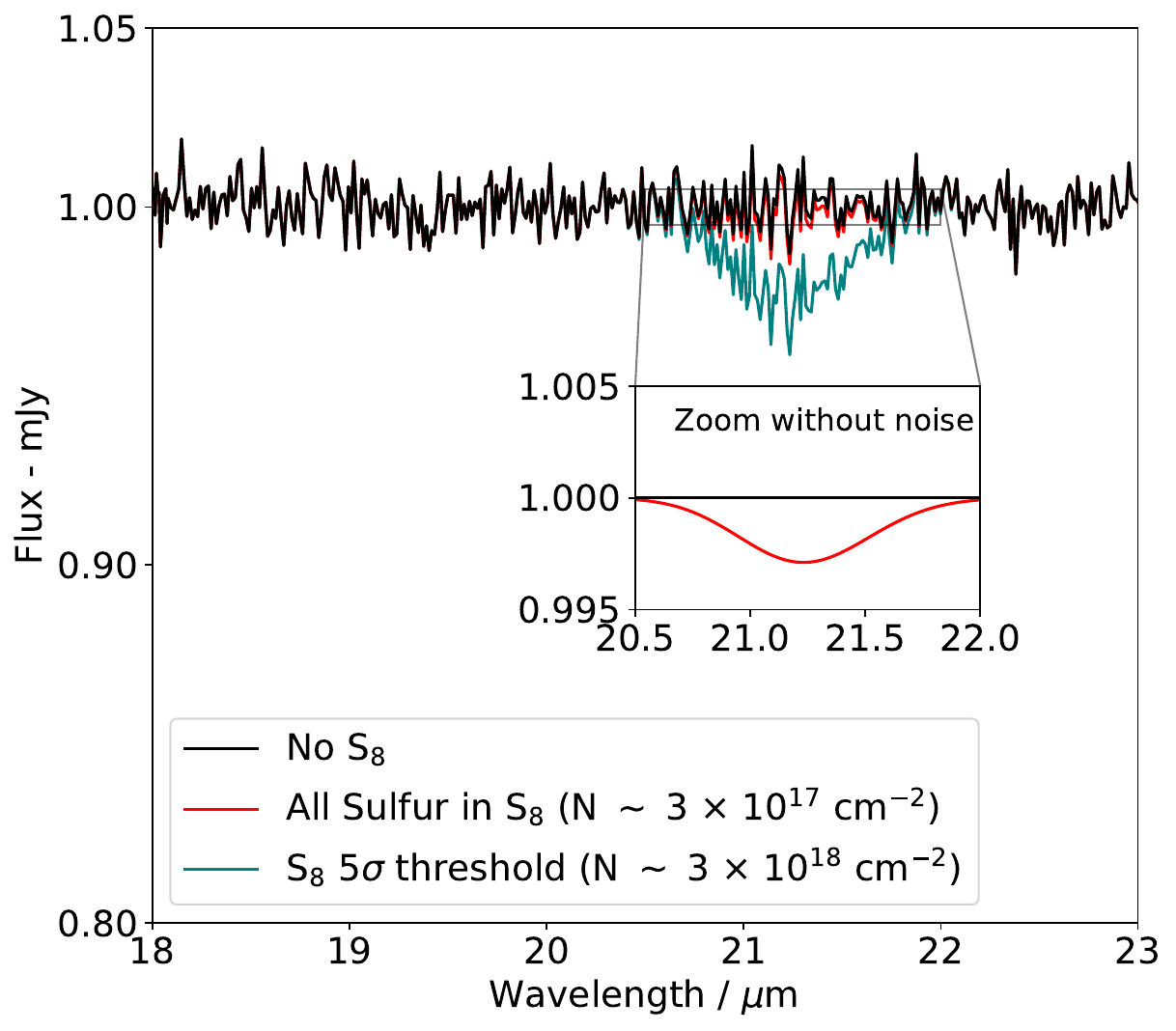}
    \includegraphics[width=0.99\linewidth]{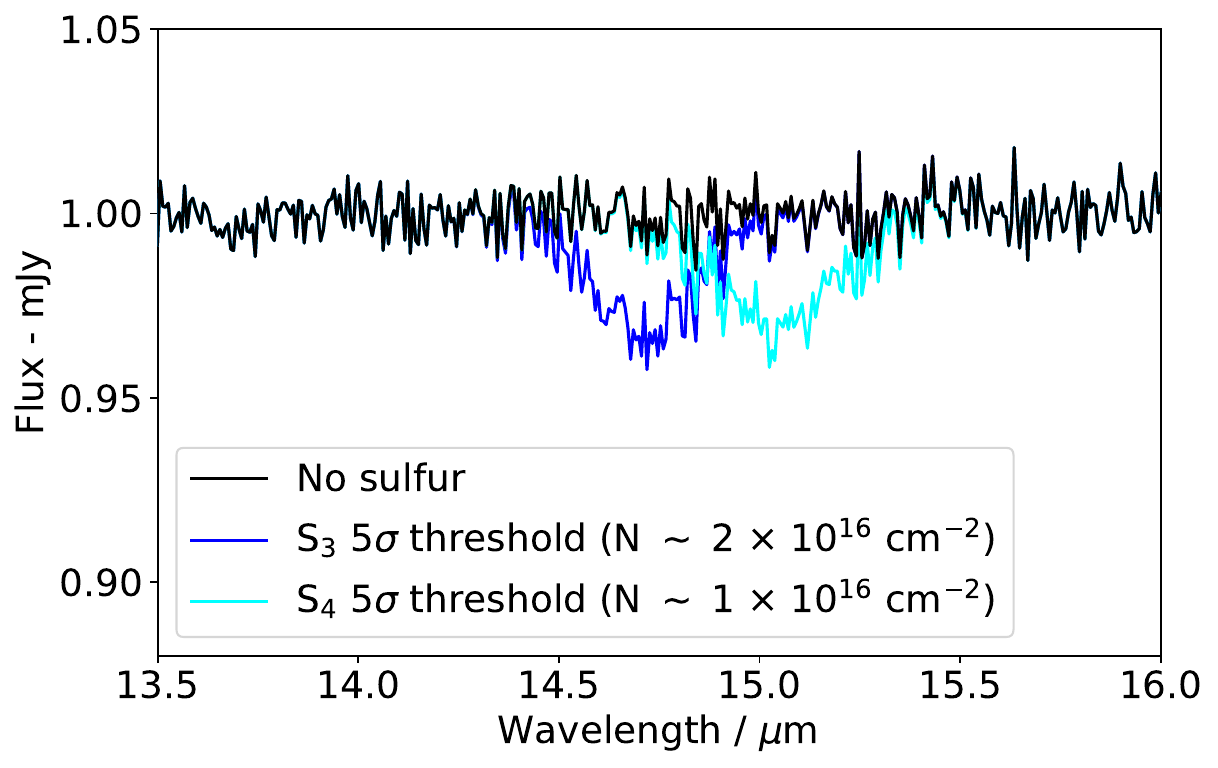}
    \caption{Sulphur allotropes detection thresholds. Top: S$_8$ 5$\sigma$ threshold estimation in our synthetic spectra generator. The spectrum in black contains no sulphur, while in red, where we consider that the entire cosmic sulphur abundance is locked in S$_8$ (see text for calculation), it still does not result in a possible detection. In teal blue, the column density needed to detect S$_8$ with a 5$\sigma$ threshold is an order of magnitude greater than the cosmic sulphur abundance.
    Bottom: Similarly to upper panel, but for the S$_3$ and S$_4$ 5$\sigma$ threshold estimations (in blue and cyan, respectively). The two bands are isolated in this figure but can be seen overlapped by other features in Fig.~\ref{fig:miri20_lab}.} 
    \label{fig:S8_threshold}
\end{figure}

Observations of translucent clouds have revealed that up to $\sim$90\% of the S atoms in diffuse clouds \citep[assuming S/H $\sim$ 1.5 $\times$ 10$^{-5}$ in diffuse cloud mainly in the form of S$^{+}$,][]{jenkins_unified_2009} deplete onto the surface of negatively charged grains prior to the formation of dense clouds \citep{2019A&A...624A.105F}. This depleted sulphur could lead to the formation of S-chains on the surface of dust grains \citep{2022AA...657A.100C}.
In comet 67P, close to 25\% of the detected volatile sulphur is in atomic form \citep{calmonte_sulphur-bearing_2016}.
Its origin is not clear, but one hypothesis is that it could be due to fragmentation of semi-refractory sulphur allotropes.
Indeed, previous laboratory and theoretical work had already suggested that sulphur allotropes (mainly S$_8$) could harbour a significant fraction of the missing sulphur in the dense ISM \citep{2014MNRAS.443..343J,2022AA...657A.100C}.

The presence of small sulphur chains in the comet could also be the result of UV processing of H$_2$S-bearing ices, as evidenced in laboratory experiments \citep{2022AA...657A.100C}. In these experiments, in addition to S-chains, other S-bearing molecules such as H$_2$S$_2$ or the radical HS$_2^.$ are formed. These co-products could also be present in interstellar ice mantles. 
The latter species has recently been detected in the gas-phase of a photon-dominated region (PDR) \citep{2017ApJ...851L..49F}.

Allotropes as the S-bearing reservoir of the missing sulphur is a hypothesis that was proposed by \citet{1997ApJ...479..839P} and further explored in \citet{2010A&A...509A..67G,jimenez-escobar_sulfur_2011}, with more recent agreements presented in models by \citet{2020ApJ...888...52S}. However, due to their low infrared band strengths (see Sect.~\ref{lab}), observing the (semi-)refractory S-bearing molecules seems to be challenging. 
The case of S$_8$ is particularly adverse due to the lower efficiency in the MIRI MRS Channel 4 than in other detector channels. The strongest S$_8$ IR feature falls in Channel 4, where the efficiency has dropped significantly during flight, but seems to have stabilised now with a photon conversion efficiency ranging from $\sim$ 0.015 to 0.03 in the Channel 4 MEDIUM, where S$_8$ is located \citep{Gillian_2023,Argyriou_2023,Gasman_2023}. The synthetic spectra do not include filter throughput curves as of now.
According to the calculated S$_8$ IR band strength, this feature seems highly unlikely to be observed by the JWST, even if all S atoms were locked in S$_8$ molecules. By considering the cosmic abundance of sulphur \citep[(S/H) = 1.5 $\times$ 10$^{-5}$,][]{jenkins_unified_2009}, divided between the eight atoms of sulphur (X(S$_8$) = 1.5 $\times$ 10$^{-5}$ / 8), we converted it to a column density using the following conversion equation: N$_{\rm x}$ = X(S$_8$) $\times$ N${\rm_H}$ $\times$ A$_V$, with N${\rm_H}$ = 1.9 $\times 10^{21}$ molecules cm$^{-2}$ and $A_{\rm V}$ = 40. This leads to N$_{S_8}$ = 3.04 $\times 10^{17}$ molecules cm$^{-2}$. In Fig.~\ref{fig:S8_threshold}, we show how this would appear on an IR spectrum, plotting the spectrum without S$_8$ in black, with the column densities computed here (barely visible) in red, and the detection threshold (3 $\times 10^{18}$ molecules cm$^{-2}$) that would be needed to detect S$_8$ in teal. The column density of the threshold is an impossible scenario, first in the quantity of sulphur required along the line of sight (A$_V$ > 400 mag and all the sulphur locked in S$_8$) and secondly because the time required to lock it all in ices would be infinitely long.

In Fig.~\ref{fig:S8_threshold}, we also tried to reproduce the 5$\sigma$ detection thresholds for S$_3$ (in blue) and S$_4$ (cyan) with the band strength values we computed earlier and applying the bandwidth from S$_8$ since we don't have measurements for the smaller allotropes. For the sake of clarity, this is on a spectrum without any other features from different species present in the area (we would encounter features from H$_2$O, CH$_3$OH, and CO$_2$) but we included the two bands in Appendix~\ref{fig:miri20_lab}.
With these approximations, the detection threshold could be reached with column densities of 2 $\times$ 10$^{16}$ and 1 $\times$ 10$^{16}$ molecules cm$^{-2}$, respectively.
S$_3$ was part of the species studied by \citet{2022AA...657A.100C} but was not detected during the heating of the ice samples, seemingly hinting that the sulphur allotropes might favoured the dimerisation of S$_2$ molecules, where allotropes with an even number of S atoms are less formed. S$_4$, on the other hand, was detected and is favourably produced.
We stress that the column densities we have derived are in the `perfect' scenario where the subtraction of the water feature at 760 cm$^{-1}$, CH$_3$OH feature at 700 cm$^{-1}$ and CO$_2$ feature at 653 cm$^{-1}$) was done correctly. The identification of the two bands in the presence of the other species in the ice could be very difficult.
According to our calculations, these smaller S allotropes could be detected if they contain about 20\% to 10\% of the sulphur budget but it should be remembered that the band strengths have been estimated from theoretical calculations and that we do not have any available data for the bandwidths.
It would be desirable to have laboratory experiments of both pure ices and mixtures in water ice that could help in providing more accurate values of thresholds.

\subsection{Comparison with chemical models}\label{sec:comp_model}

From all the cases discussed here, the easiest one to reproduce and compare to specific observations would be the dense cloud environment.
For this purpose, we also used the Nautilus abundance predictions for TMC-1 from \citet{navarro-almaida_gas_2020} to create a new synthetic spectrum, that is presented in Appendix~\ref{appendixb}.
The TMC low-mass star forming region is of particular interest, lying at a relatively close distance of 140 pc \citep{2002ApJ...575..950O}, where many studies have been conducted to understand the physics and chemistry of the cloud \citep{1995ApJ...445L.161M,2008ApJ...680..428G,2013ChRv..113.8710A,2019A&A...630A.136N}. The prominent filament TMC-1 has been observed thoroughly, showing a peculiar chemistry rich in carbon chains \citep{2007ApJ...664L..43B,2020A&A...641L...9C,2022ApJ...924...21S}. \citet{2013ChRv..113.8710A} has shown that its high C/O ratio in the dense gas is a consequence of the depletion of water molecules at the surface of grains, forming thick ice mantles. 
The GEMS IRAM 30m Large Program \citep{2019A&A...624A.105F} conducted an extensive study of the gas phase. Later, the study of \citet{2023A&A...670A.114F} aimed specifically at quantifying sulphur species, showing that the sulphur depletion factor was of $\sim$ 20 to $\sim$ 100. The filament exhibits particular physical conditions, both moderate UV field and high cosmic ray flux, which are the most favourable to form large sulphur chains on grain surfaces \citep{2022AA...657A.100C}.

\begin{figure}
    \centering
    \includegraphics[width=0.99\linewidth]{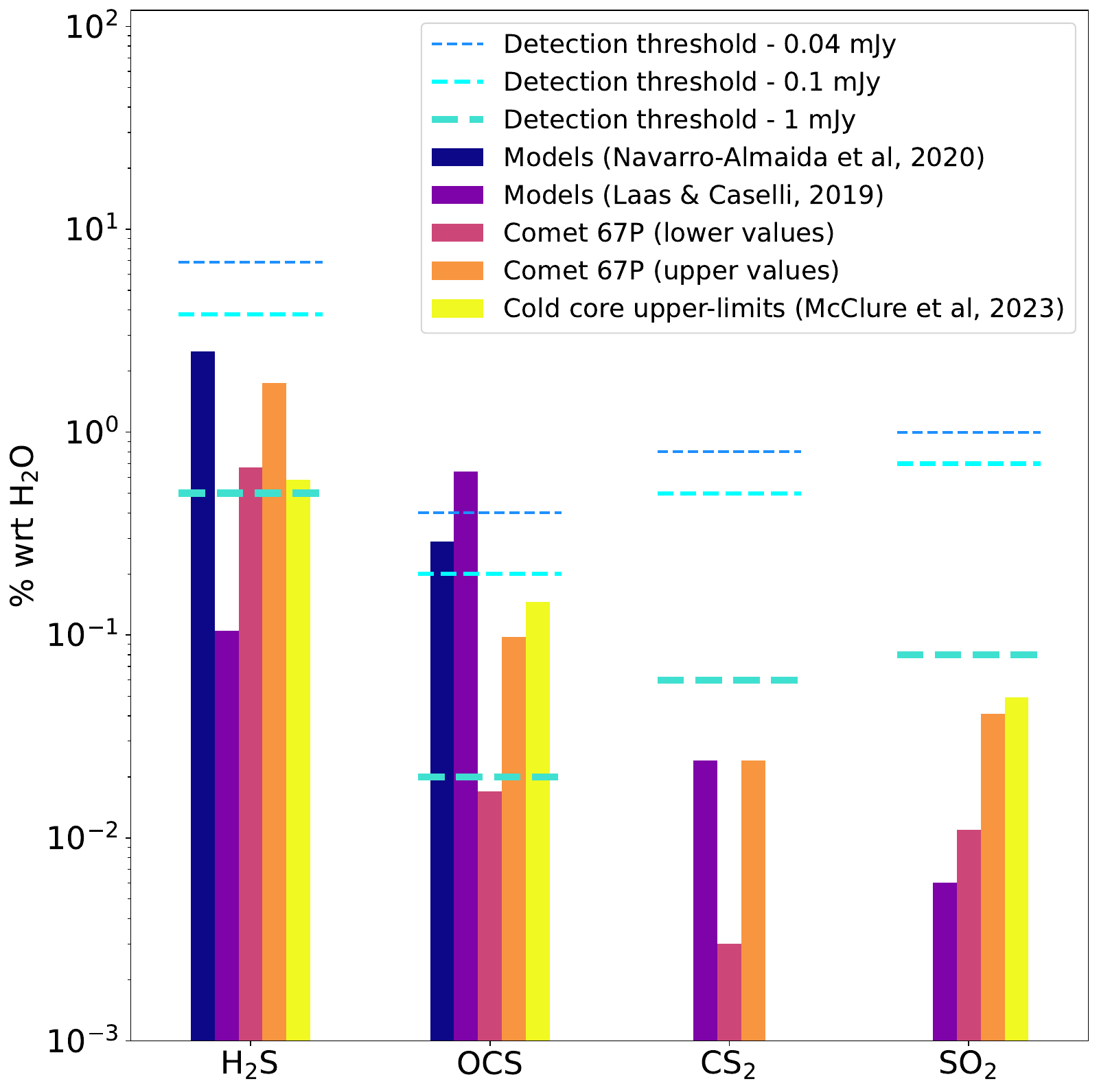}
    \caption{Comparison between abundances with respect to water predicted by models from \citet{navarro-almaida_gas_2020} and \citet{2019A&A...624A.108L} (Stage 3, density of n$\rm_H$ = 10$^{5}$ cm$^{-3}$), cometary observations \citet{calmonte_sulphur-bearing_2016}, dense cloud observations from \citet{2023NatAs...7..431M} with, in dashed blue lines, the detection thresholds we determined for the dense cloud environment. Note that SO$_2$ in dense clouds is expected to be less abundant than in YSOs and MYSOs, where it could be detected. } 
    \label{fig:S_comparison_cold_core}
\end{figure}

One of the main results from \citet{navarro-almaida_gas_2020} was predicting H$_2$S to be the main sulphur reservoir (with 2.5\% abundance relative to water) in the TMC-1 filament. Only H$_2$S and OCS could be possibly detected, in bright (1 mJy) and faint (0.1 mJy) flux in our synthetic spectra (see Fig.~\ref{fig:bg_star_no_continuum_sis_nautilus}), while CS$_2$ is not formed efficiently in Nautilus by any pathway (despite the formation routes being included in the chemical network). This adds more doubts with respect to the possible detectability of CS$_2$ in dense clouds, despite the thresholds values being low compared to the abundance of water. For this particular case, SO$_2$ is not predicted to be formed efficiently (0.00003\% with respect to water); but within the same studies, using different parameters to reproduce Barnard 1-b, it reaches 0.08\% with respect to water. The initial abundance of sulphur used in these models is 1.5 $\times$ 10$^{-5}$, meaning that more than 95\% of the sulphur is locked in ices.

The other similar model predictions by \citet{2019A&A...624A.108L} can also be compared with our detection thresholds. H$_2$S is within a factor $\sim$ 5 in the brightest source while OCS would be solidly detected at all thresholds. However CS$_2$ and SO$_2$ are lower by a factor $\sim$ 8 and $>$ 12. The initial sulphur abundance used there is of 1.66 $\times$ 10$^{-5}$ \citep{Esteban_2004} and where more than 99\% of this abundance is locked in ices by the end of the simulations.
The main differences between this model and the previous one include the diffusion coefficients used, the enhanced accretion of atomic S$^+$ and that chemical desorption is disabled for all species. One major consequence is that H$_2$S is not the main reservoir in most cases. Moreover, the formation of CS$_2$ is enhanced, but still its abundance is too low to be detectable. Both models are within the range of H$_2$S abundances observed in comets but only the model by \citet{2019A&A...624A.108L} is consistent with the H$_2$S upper limit reported by \citet{2023NatAs...7..431M} for the dense cloud Chamaeleon I. However, the OCS abundance reported by \citet{2023NatAs...7..431M}, as well as cometary OCS abundances, are more consistent with the predictions from \citet{navarro-almaida_gas_2020}.

Figure~\ref{fig:S_comparison_cold_core} summarises the different model predictions, comets, and cold core observations for the four different molecules in our sample. 
Our calculated detection limits for the dense cloud environment are shown as dashed blue lines for each molecule. Given the solar abundance of atomic sulphur, the non-detection of these species would imply that none of them is the main sulphur reservoir in the ice. This is particularly important for H$_2$S, which is expected to be one of the most abundant icy sulphur species. Furthermore, the derived H$_2$S/OCS and SO$_2$/OCS abundance ratios (or their upper bounds in the case of non-detection) would be extremely useful to test our current knowledge of sulphur surface chemistry and to constrain our chemical models, a mandatory step for any future research. Unfortunately, we could not determine the amount of sulphur in the volatiles with these measurements since it is possible that the sulphur is distributed among several species in amounts that may be just below the detection threshold. Alternatively, most of the sulphur may be trapped in sulphur allotropes and would not be detectable with our observations (as we mention in Sect.~\ref{allotropes}). Being aware of this limitation, we would like to emphasise that the detection of any new S carrier, in addition to OCS, would have a strong impact on the understanding of sulphur chemistry and even change the paradigm of the current view of sulphur chemistry.

\section{Conclusions}

In this paper, we study the detectability with JWST of five S-bearing species that have been identified as potential sulphur reservoirs in the solid-phase: H$_2$S, OCS, SO$_2$, CS$_2$, and S$_8$. We compiled different IR band strengths, positions, and widths of the targeted sulphur species from literature. As some values were not available for our study, we measured new band parameters for CS$_2$ and S$_8$ that were used for subsequent modelling.
Using the synthetic ice spectrum generator SynthIceSpec, we identified which S-bearing species' bands would overlap with absorption features of the major ice components (H$_2$O, CO, CO$_2$ and CH$_3$OH), considering a mixed ice composition. Unlike OCS and SO$_2$, H$_2$S and CS$_2$ are close to strong absorption bands making them harder to identify.
In addition, we were able to compute threshold values to allow for a 5$\sigma$ detection of four sulphur species in three different environments (dense cloud, LYSO, and MYSO) at three different constant fluxes (1, 0.1, and 0.04 mJy). The rms used for this study was derived from \citet{2023NatAs...7..431M} and directly impacts our estimations. The detection thresholds allowed us to determine the specific scenarios where these values could be expected to be realistic. 
We summarise our work in the following points:

\begin{itemize}  
    \item For the H$_2$S feature at 3.92 $\mu$m, detection is heavily affected by the ice composition: the presence of methanol (combination mode) causes an overlap between the two bands. Methanol is expected in large quantities in all three environments, meaning that it is most certainly expected to challenge H$_2$S detection. Furthermore, dust scattering would affect the band area, with stronger red features observed in the water band, overlapping with our targeted feature. With our synthetic ice spectra, we would expect a detection feasible for an abundance range of 0.5\%-6.9\% (dense clouds), 1.1\%-17.5\% (LYSO) and 1.0\%-15.5\% (MYSO) with respect to water. 

    \item The OCS feature at 4.9 $\mu$m has been tentatively detected in different regions \citep[respectively $\le$ 0.22\%, $\le$1.6\% and 0.03–0.16\% with respect to water in dense clouds, class 0, and high-mass protostars,][]{2015ARA&A..53..541B}.
    With our synthetic spectra, the detection thresholds for OCS would range from {0.02\%-0.4\% (dense cloud), 0.05\%-1.0\% (LYSO), and 0.05\%-1.2\% (MYSO)} with respect to H$_2$O. These values are both within the upper-limits reported in the literature and within model predictions. 
    
    \item The CS$_2$ feature at 6.7 $\mu$m is located in a spectral region where both water and methanol features are expected to be strong. Its shape can be deformed when mixed with water, which, in the presence of other molecules, could be shifted as well. It has not been reported from observations so far but it has the highest band strength measured in our four targeted molecules. The threshold ranges we expect are 0.06\%-0.8\% (dense cloud), 0.1\%-1.0\% (LYSO), and 0.08\%-1.6\% (MYSO) with respect to water. 
    
    \item The SO$_2$ feature at 7.5 $\mu$m is highly dependent on the ice composition, as it is located in a very `populated' spectral region, including bands of COMs and possible overlapping effect from 
    more abundant species (e.g. OCN$^-$). The threshold ranges we computed are 0.08\%-1.0\%, 0.15\%-2.1\%, and 0.12\%-1.7\% with respect to water in a dense cloud environment, a LYSO, and a MYSO, respectively. Most of our thresholds are in agree with past upper-limit estimations and predictions.

    \item S$_8$ has been identified to be a possible reservoir for solid sulphur. With our first reported band strength measurement, we propose that its detection is very unlikely. Even by considering that the entire cosmic abundance of atomic S is locked in this form, it would still be impossible to definitively detect this molecule. 

    \item Considering the thresholds computed here, the existence of a main sulphur reservoir between the four targeted species seems unlikely, as it would already have been detected. It is possible that sulphur is well divided between these species or that most of it will remain invisible from our IR studies with the JWST.
    
\end{itemize}

Overall, the detection of S-bearing species in ices could be very challenging in most cases. 
In certain specific environments, the S-bearing species we study in this work could be favourably formed.  Therefore, it would be detectable, as the column density thresholds we derived are not particularly high, however, they are extremely dependent on the ice composition, dust properties, and scattering. We have not considered important data extraction and treatment (e.g. continuum subtraction, stellar photosphere...) that could affect the derived thresholds either.
JWST observations cannot provide information on the total amount of sulphur in ices if the majority of sulphur were to be locked in species with low band strengths and thereby remaining hindered with respect to our observations. Nonetheless, JWST could provide the first detections of (or stringent upper limits on) H$_2$S and SO$_2$ column densities, which which are essential for astrochemistry since this would provide the first possible benchmark of our chemical models with sulphur ice observations. This comparison will help us resolve the sulphur problem and will open new paths for future research.
With this study, we hope to provide a pathway to help observers to identify possible features as S-bearing molecules in their studies. 

The thresholds published in this paper have been calculated for four representative cases of low-mass and high-mass star forming regions and a given rms. It should be noted that they cannot be extrapolated to all astronomical objects. Different continuum flux values, chemical abundances, and/or observational settings could lead to different values. To ensure accurate thresholds, the calculations need to be repeated for each particular case.

\begin{acknowledgements} 
This project has received funding from the European Research Council (ERC) under the European Union’s Horizon Europe research and innovation programme ERC-AdG-2022 (GA No. 101096293).
A.S.-M.\ acknowledges support from the RyC2021-032892-I grant funded by MCIN/AEI/10.13039/501100011033 and by the European Union `Next GenerationEU’/PRTR, as well as the program Unidad de Excelencia María de Maeztu CEX2020-001058-M, and support from the PID2020-117710GB-I00 (MCI-AEI-FEDER, UE). 
H.C. and G.M.M.C. were funded by project PID2020-118974GB-C21 of the Spanish Ministry of Science and Innovation. 
R.M.-D. was supported by a La Caixa Junior Leader grant under agreement LCF/BQ/PI22/11910030. 
D.N.-A. acknowledges funding support from Fundaci\'on Ram\'on Areces through its international postdoc grant program. 
AF also thanks project PID2023-146675NB-I00 funded by the Spanish Ministry of Science and Innovation/State Agency of Research MCIN/AEI/ 10.13039/501100011033 and by “ERDF A way of making Europe”. 
J.A.N. and E.D acknowledge support from the French program ``Physique et Chimie du Milieu
Interstellaire'' (PCMI) of the CNRS/INSU with the INC/INP cofunded by the CEA and CNES. B.E. acknowledges support by grant PTA2020-018247-I by the Spanish Ministry of Science and Innovation/State Agency of Research MCIN/AEI.

\end{acknowledgements}

\bibliography{aa52900-24corr}
\bibliographystyle{aa}

\newpage
\appendix

\onecolumn
\section{Additional material}

\begin{table*}[!h] 
\centering
\caption{Gaussian parameters used in SynthIceSpec}
\begin{tabular}{|l|cccccccc|}
\hline
Molecule & Position & Position & Gaussian intensity     & FWHM & Mode      & Composition  & Reference      & \\  
\multicolumn{1}{|c|}{} & \multicolumn{1}{c}{(cm$^{-1}$)} & \multicolumn{1}{c}{($\mu$m)} &  \multicolumn{1}{c}{(cm/molecule)} & \multicolumn{1}{c}{(cm$^{-1}$)} &  \multicolumn{3}{c}{}                     &   \\

\hline
\multicolumn{8}{|c}{Major ice components}               & \\
\hline
H$_2$O          & 3702.6  & 2.7 & 6.7 $\times$ 10$^{-20}$  & 11.3  & Dangling OH        & H$_2$O:CH$_3$OH:CO$_2$ (9:1:2)   & $a$& \\  
H$_2$O          & 3654.3  & 2.7 & 8.9 $\times$ 10$^{-19}$  & 91.1  & Dangling OH        & H$_2$O:CH$_3$OH:CO$_2$ (9:1:2)   & $a$ &\\  
H$_2$O          & 3442.7  & 2.9 & 1.3 $\times$ 10$^{-17}$  & 179.1 & O-H stretch        & H$_2$O:CH$_3$OH:CO$_2$ (9:1:2)   & $a$& \\  
H$_2$O          & 3283.9  & 3.0 & 1.1 $\times$ 10$^{-17}$  & 227.6 & O-H stretch        & H$_2$O:CH$_3$OH:CO$_2$ (9:1:2)   & $a$& \\  
H$_2$O          & 3273.9  & 3.0 & 6.0 $\times$ 10$^{-17}$  & 298.5 & O-H stretch        & H$_2$O:CH$_3$OH:CO$_2$ (9:1:2)   & $a$& \\  
H$_2$O          & 1655.0  & 6.0 & 7.8 $\times$ 10$^{-17}$  & 160.0 & Libration          & H$_2$O:CS$_2$ (7:1)              & $b$& \\  
CO              & 2137.5  & 4.6 & 1.1 $\times$ 10$^{-17}$  & 7.9 & C-O stretch        & H$_2$O:CO (4:1)                  &$c$&\\ 
CO$_2$          & 2350.0  & 4.2 & 2.8 $\times$ 10$^{-17}$  & 11.4 & C-O stretch        & H$_2$O:CH$_3$OH:CO$_2$ (9:1:2)   & $a$& \\  
CO$_2$          & 2345.0  & 4.2 & 3.1 $\times$ 10$^{-17}$  & 16.4 & C-O stretch        & H$_2$O:CH$_3$OH:CO$_2$ (9:1:2)   & $a$& \\  
CH$_3$OH        & 3250.0  & 3.0 & 1.1 $\times$ 10$^{-16}$  & 235.0 & C-H stretch        & H$_2$O:CH$_3$OH:CO$_2$ (9:1:2)   & $a$& \\   
CH$_3$OH        & 3005.0  & 3.3 & 1.1 $\times$ 10$^{-17}$  & 100.0 & C-H stretch        & H$_2$O:CH$_3$OH:CO$_2$ (9:1:2)   & $a$ &\\  
CH$_3$OH        & 2833.5  & 3.5 & 6.0 $\times$ 10$^{-18}$  & 38.9 & C-H stretch        & H$_2$O:CH$_3$OH:CO$_2$ (9:1:2)   & $a$& \\   
CH$_3$OH        & 2610.0  & 3.8 & 0.6 $\times$ 10$^{-18}$  & 35.2 & Combination        & H$_2$O:CH$_3$OH:CO$_2$ (9:1:2)   & $a$& \\  
CH$_3$OH        & 2530.0  & 3.9 & 0.8 $\times$ 10$^{-16}$  & 141.0 & Combination        & H$_2$O:CH$_3$OH:CO$_2$ (9:1:2)   & $a$& \\   

\hline
\hline
\multicolumn{8}{|c}{Additional relevant ice components}   &             \\
\hline
CH$_4$          &  1303.0 & 7.6 & 6.0 $\times$ 10$^{-18}$  & 9.4 &  C-H bending     & H$_2$O:CH$_3$ (20:1) & $d$   &  \\
NH$_3$          &  1642.0 & 6.0 & 2.9  $\times$ 10$^{-18}$ & 58.7 & N-H bending     & H$_2$O:NH$_3$ (15:1)        &  $e$& \\ 
NH$_3$          &  1693.0 & 5.9 & 2.8  $\times$ 10$^{-18}$ & 63.4 & N-H bending     & H$_2$O:NH$_3$ (15:1)        &  $e$ &\\ 
HCOOH           &  1708.0 & 5.8 & 2.7 $\times$ 10$^{-17}$  & 38.0 & C=O stretching  & H$_2$O:HCOOH (91:9)     & $f$& \\
HCOOH           &  1636.0 & 6.1 & 2.7 $\times$ 10$^{-17}$  & 70.0 & Combination     & H$_2$O:HCOOH (91:9)                     &  $f$ &\\
HCOOH           &  1391.0 & 7.1 & 0.7 $\times$ 10$^{-17}$  & 35.0 & C-H bending     & H$_2$O:HCOOH (91:9)                     &  $f$ &\\
HCOOH           &  1211.0 & 8.5 & 2.0 $\times$ 10$^{-18}$  & 64.0 & C-O stretching  & H$_2$O:HCOOH (91:9)                     &  $f$ &\\
H$_2$CO         &  1496.0 & 6.7 & 4.0 $\times$ 10$^{-18}$  & 11.0 & C-H$_2$ scissoring & H$_2$O:H$_2$CO (100:3)                  &  $g$ &\\
H$_2$CO         &  1719.0 & 5.8 & 9.6 $\times$ 10$^{-18}$  & 20.0 & C=O stretching  & H$_2$O:H$_2$CO (100:3)                  &  $g$ &\\
H$_2$CO         &  2822.0 & 3.5 & 4.1 $\times$ 10$^{-18}$  & 37.0 & C-H$_2$ stretching  & H$_2$O:H$_2$CO (100:3)                  &  $g$& \\
CH$_3$CHO       &  1717.0 & 5.8 & 3.1 $\times$ 10$^{-19}$  & 18.8 & C=O stretching  & H$_2$O:CH$_3$CHO (20:1)                 &  $h$ &\\
CH$_3$CHO       &  1350.0 & 7.4 & 8.9 $\times$ 10$^{-20}$  & 9.4 & C-H bending    & H$_2$O:CH$_3$CHO (20:1)                  &  $h$ &\\

\hline
\hline
\multicolumn{8}{|c}{sulphur species}          &      \\
\hline
H$_2$S          & 2547.7 & 3.9 & 3.7 $\times$ 10$^{-17}$  & 64.1 & H-S stretch        & H$_2$O:H$_2$S (100:7.5)             &  $i$& \\
OCS             & 2047.0 & 4.9 & 1.2 $\times$ 10$^{-16}$  & 23.2 & C-S stretch        & H$_2$O:OCS (20:1)                   & $d$&\\
CS$_2$          & 1536.7 & 6.5 & 1.1 $\times$ 10$^{-17}$  & 8.43 & C-S stretch        & H$_2$O:CS$_2$ (7:1)                & $b$ &\\
CS$_2$          & 1528.3 & 6.5 & 2.7 $\times$ 10$^{-17}$  & 19.9 & C-S stretch        & H$_2$O:CS$_2$ (7:1)                & $b$& \\
CS$_2$          & 1523.6 & 6.5 & 1.0 $\times$ 10$^{-16}$  & 45.4 & C-S stretch        & H$_2$O:CS$_2$ (7:1)                & $b$ &\\
SO$_2$          & 1326.0 & 7.5 & 2.0 $\times$ 10$^{-17}$  & 9.4 & S-O stretch        & H$_2$O:SO$_2$ (5:1)             & $j$& \\
SO$_2$          & 1331.0 & 7.5 & 1.6 $\times$ 10$^{-17}$  & 5.4 & S-O stretch        & H$_2$O:SO$_2$ (5:1)              &$j$& \\
SO$_2$          & 1149.0 & 8.7 & 7.3 $\times$ 10$^{-18}$  & 8.7 & S-O stretch        & H$_2$O:SO$_2$ (5:1)             & $j$& \\

\hline\end{tabular}
\tablefoot{Gaussian parameters used in SynthIceSpec to fit every single IR feature for each species, appearing in the spectra presented in Sect.~\ref{simple_approach}, for an ice composition of the major ice species overlapping with S-bearing species. The reference laboratory spectra used to fit each Gaussian are as follows: $a$. \citet{1999A&A...350..240E}, $b$. \citet{martin_domenech_2024_13134350}, $c$. \citet{2007A&A...476..995B}, $d$. \citet{1993ApJS...86..713H}, $e$. Private communication (E. Dartois), $f$. \citet{2007A&A...470..749B}, $g$. \citet{1993Icar..104..118S}, $h$. \citet{2018A&A...611A..35T}, $i$. \citet{jimenez-escobar_sulfur_2011}, $j$. \citet{2003JMoSt.644..151S}. 
}\label{tab:simple_approach_param}
\end{table*}

\begin{figure*}[h!]
    \centering
    \includegraphics[width=0.99\linewidth]{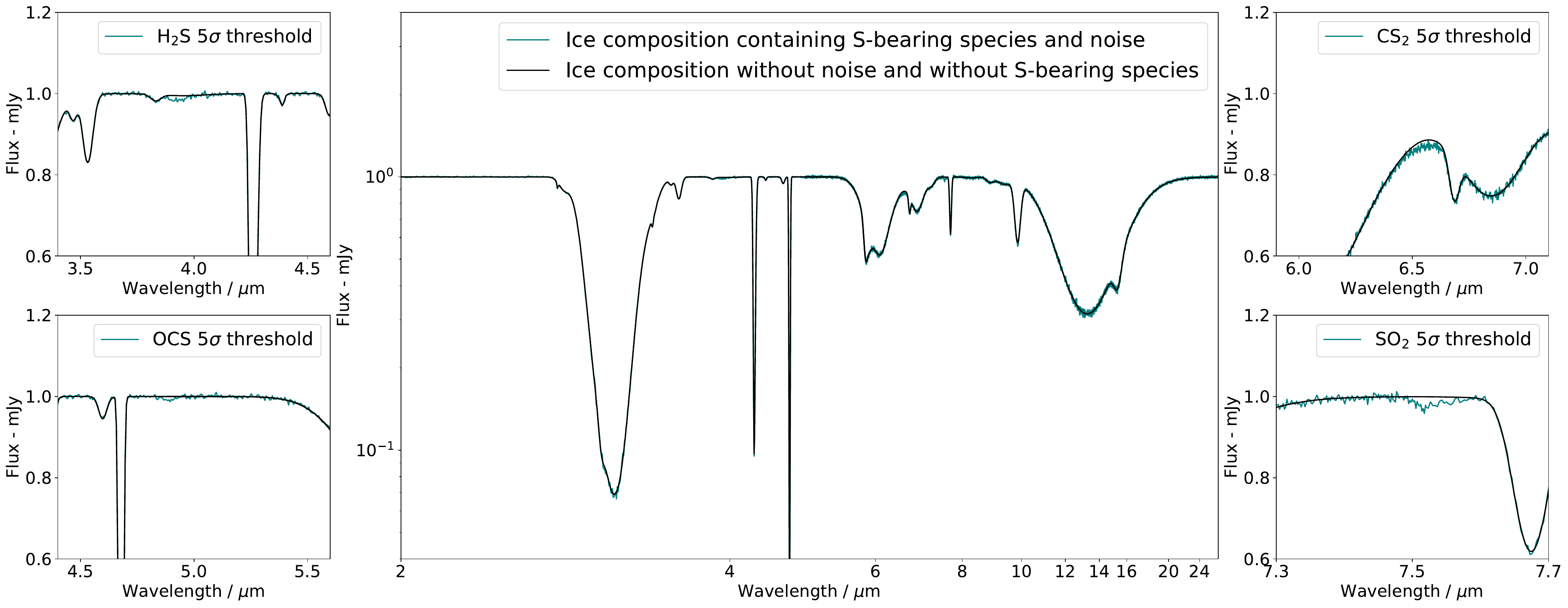}
    \includegraphics[width=0.99\linewidth]{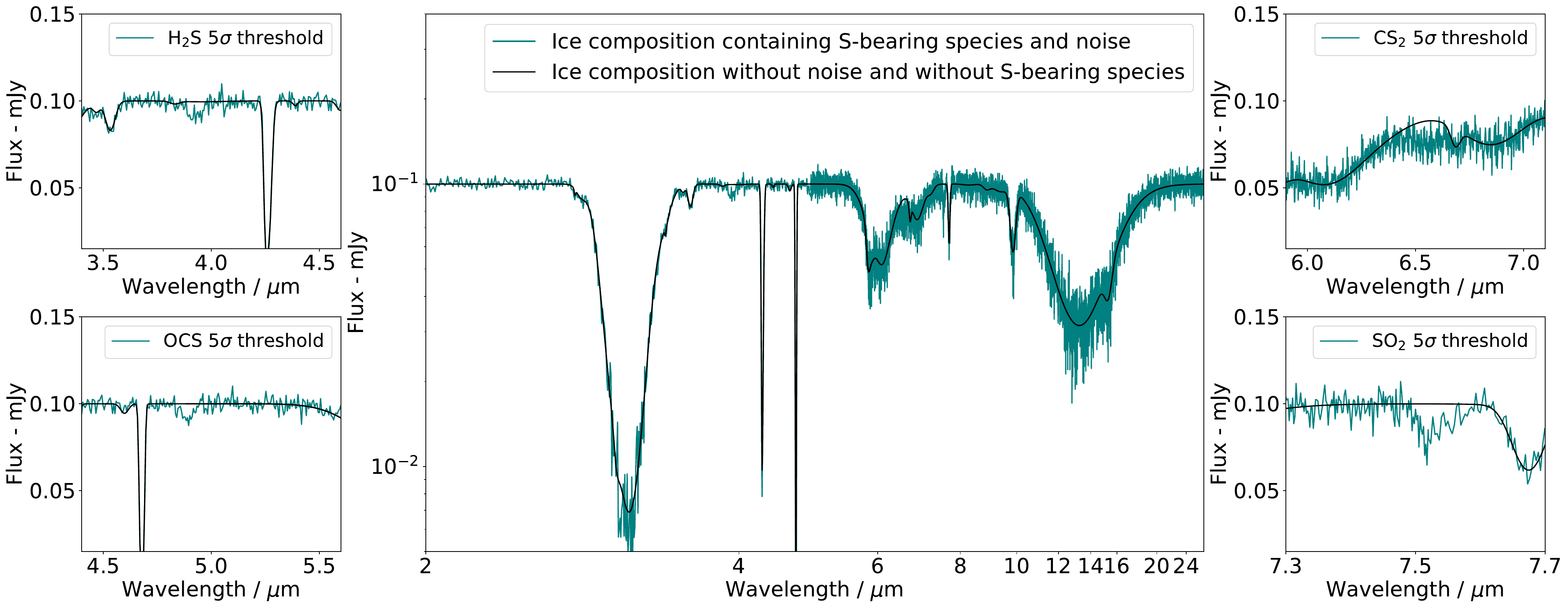}
    \includegraphics[width=0.99\linewidth]{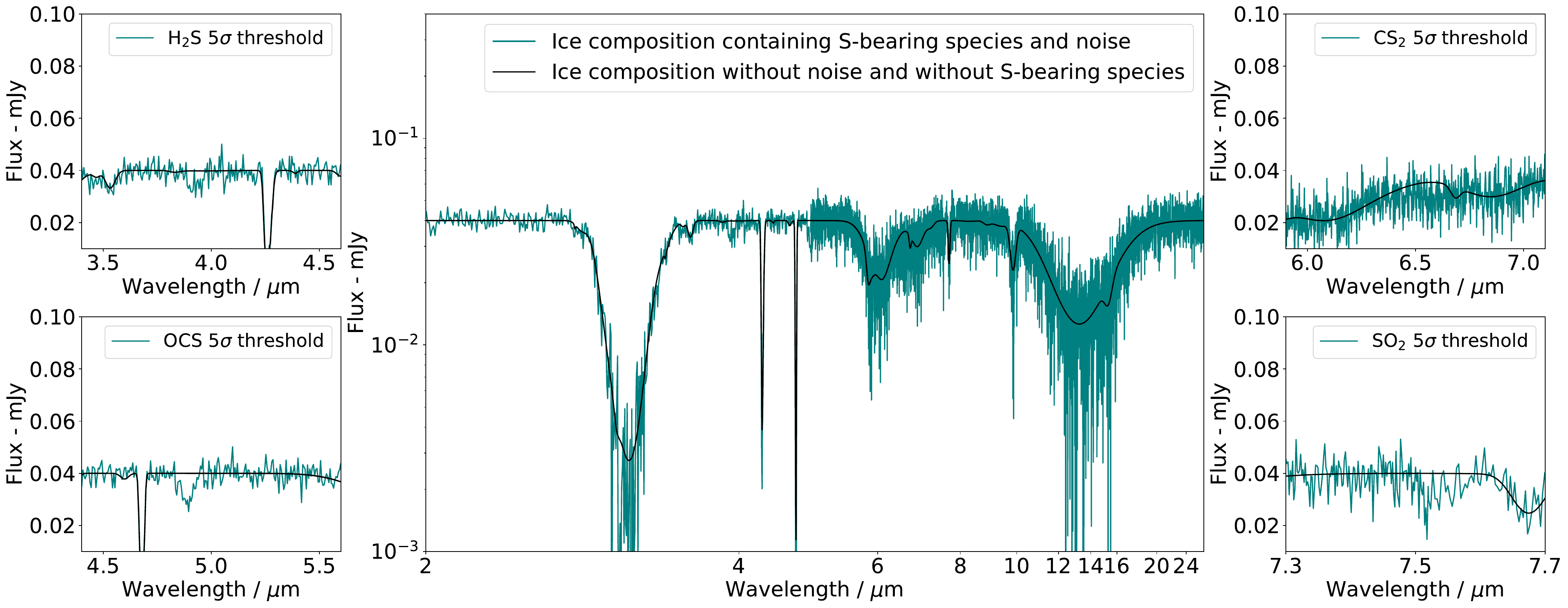}
    \caption{Threshold determination for the dense cloud environments for 3 different flat continua (1, 0.1 and 0.04 mJy), the synthetic ice spectrum in teal is the ice composition containing sulphur (and noise added), with in black the synthetic ice spectrum without sulphur and without noise. Molecular parameters used in these spectra are listed in Table~\ref{tab:simple_approach_param}. 
    Each panel shows the 5$\sigma$ threshold for one of the S-bearing species considered (on the left, top to bottom: H$_2$S, OCS; on the right, top to bottom: CS$_2$, SO$_2$). }
    \label{fig:bg_star_no_continuum_sis}
\end{figure*}

\begin{figure*}[h!]
    \centering
    \includegraphics[width=0.99\linewidth]{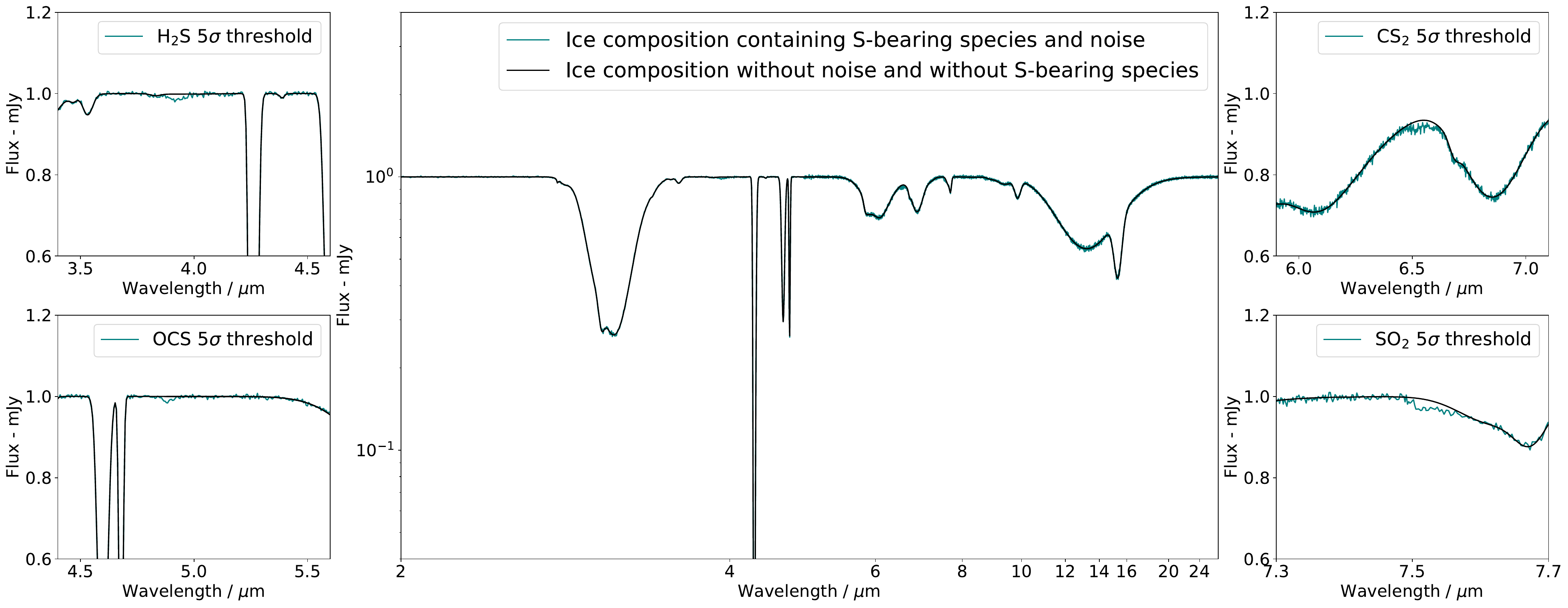}
    \includegraphics[width=0.99\linewidth]{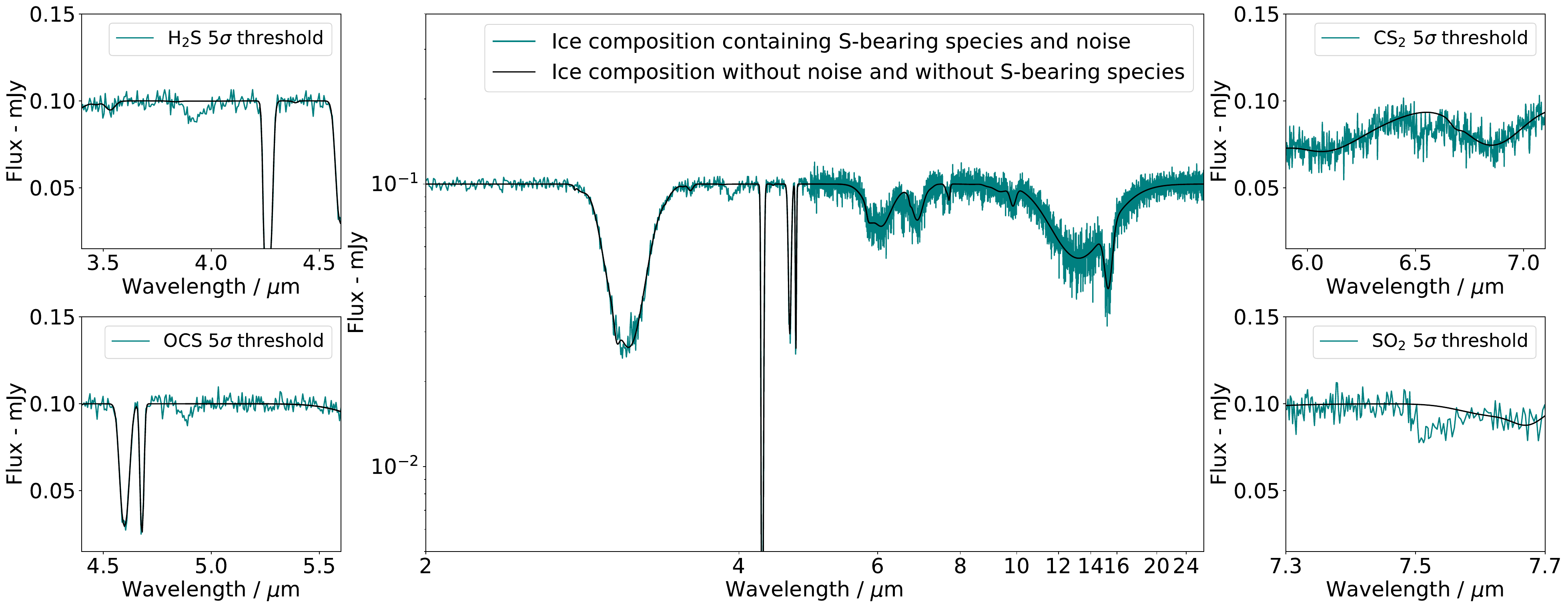}
    \includegraphics[width=0.99\linewidth]{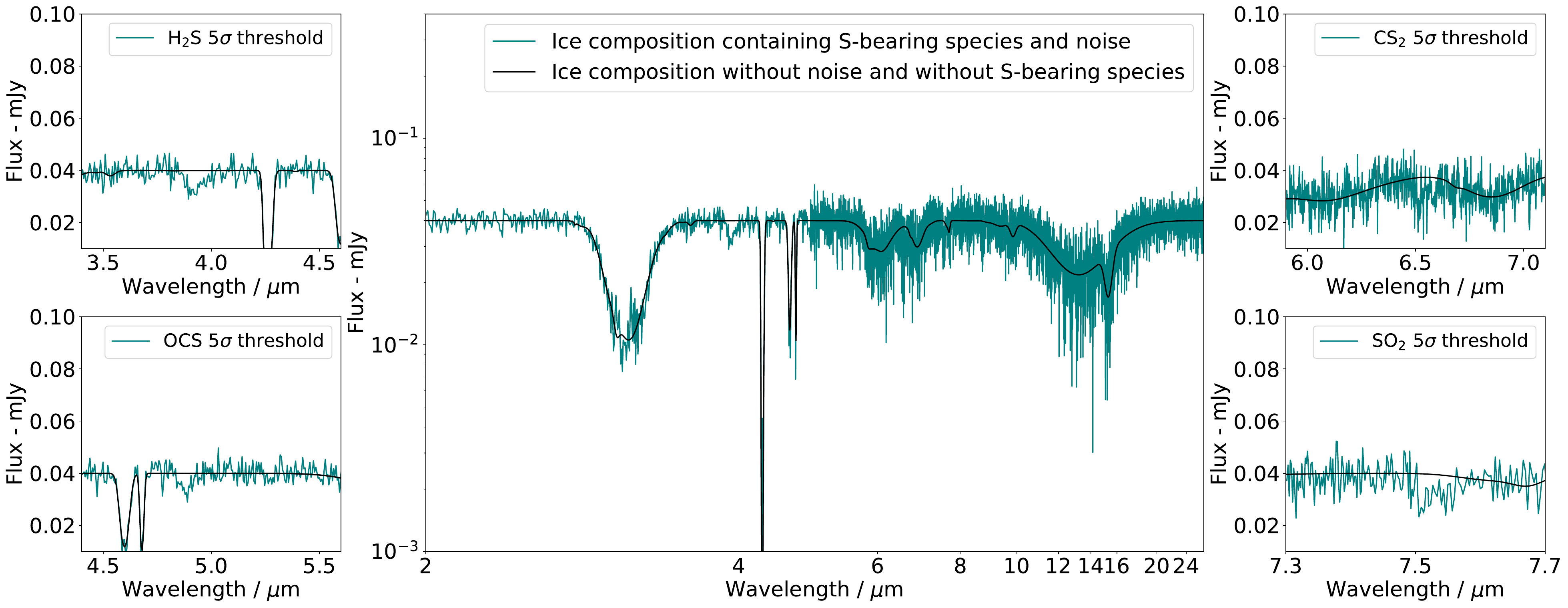}
    \caption{Same as Fig.~\ref{fig:bg_star_no_continuum_sis}, but for the LYSO environment.
    } 
    \label{fig:class0_no_continuum_sis}
\end{figure*}

\begin{figure*}[h!]
    \centering
    \includegraphics[width=0.99\linewidth]{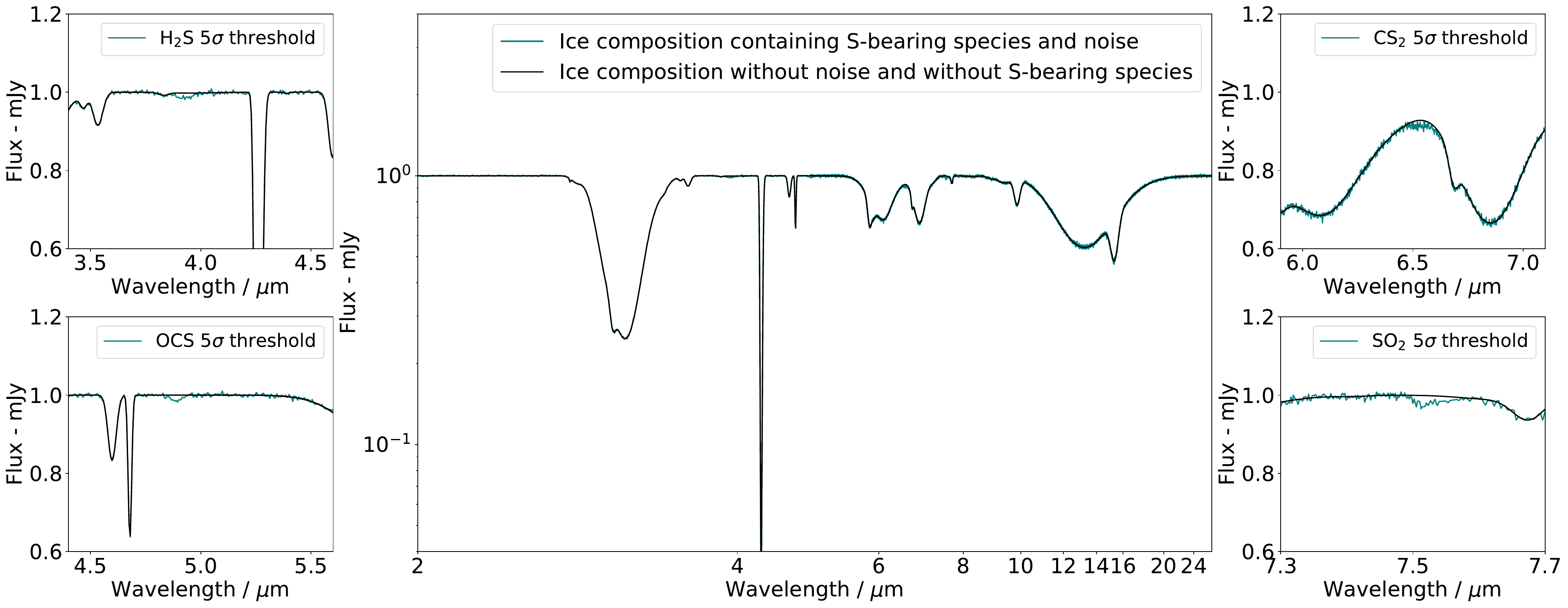}
    \includegraphics[width=0.99\linewidth]{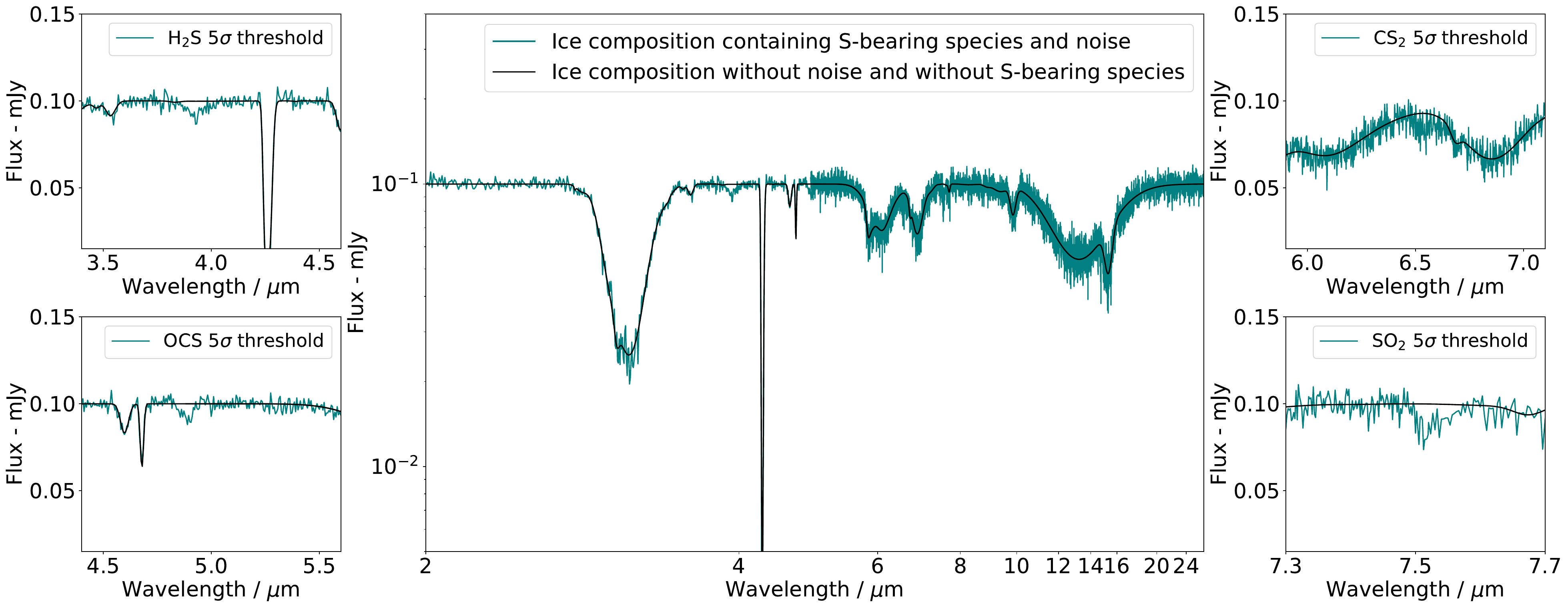}
    \includegraphics[width=0.99\linewidth]{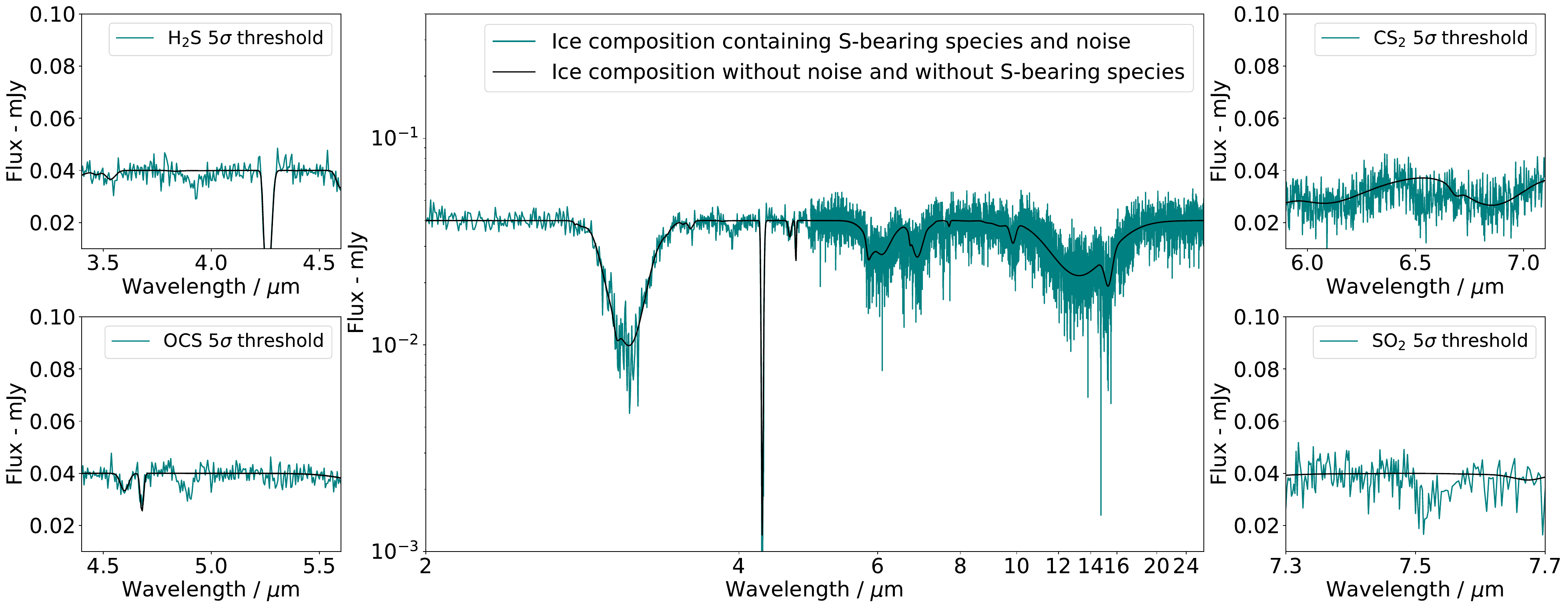}
    \caption{Same as Fig.~\ref{fig:bg_star_no_continuum_sis}, but for the MYSO environment.
    } 
    \label{fig:classI_no_continuum_sis}
\end{figure*}

\clearpage

\twocolumn

\section{Laboratory spectra in SynthIceSpec}\label{appendixa}

\begin{figure}
    \centering
    \includegraphics[width=0.97\linewidth]{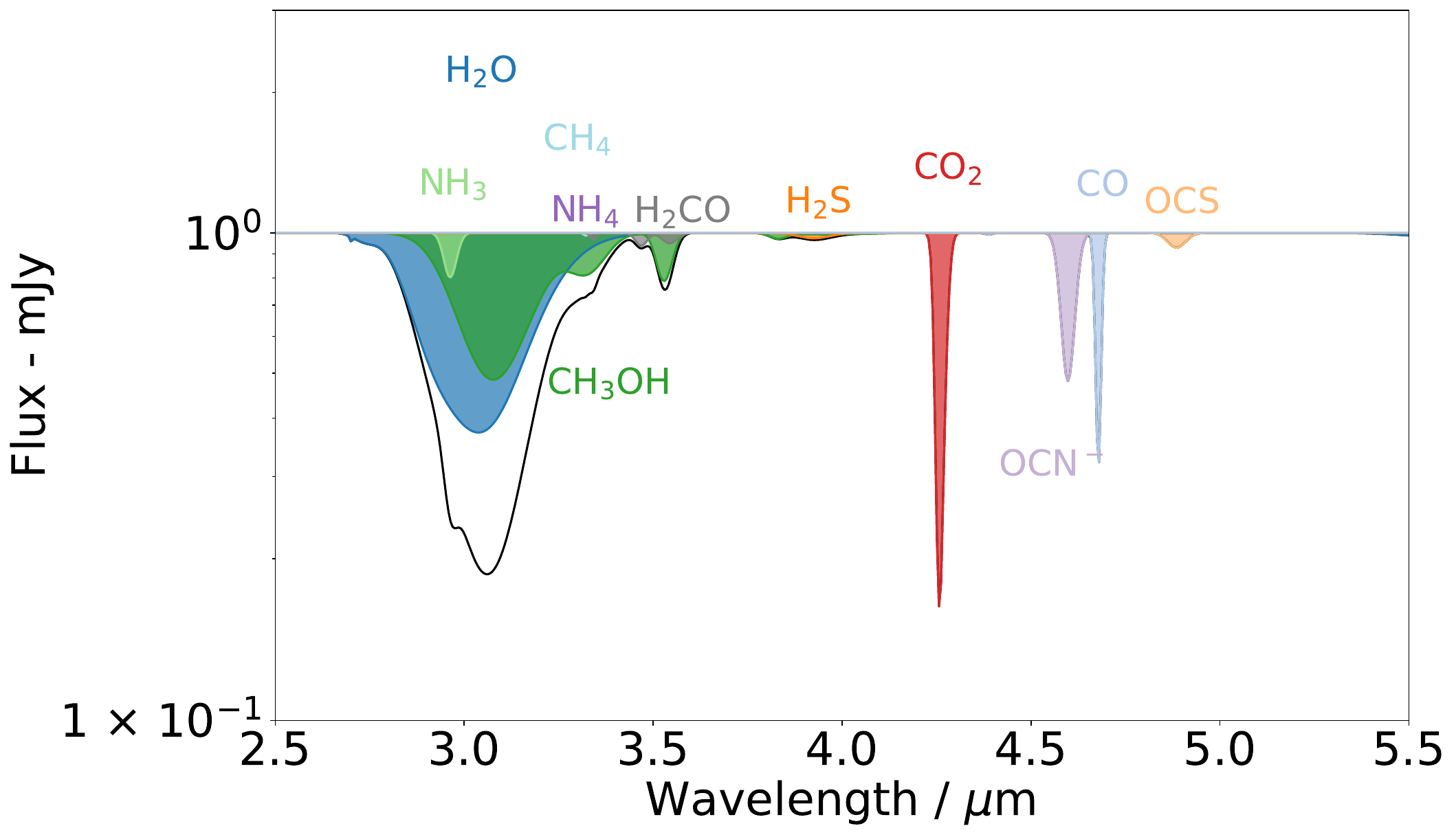}
    \caption{Different features and their contributions used in SynthIceSpec in the NIRSpec wavelength range. The overall spectrum is plotted in black. The band parameters used are those derived from Gaussian fits as listed in Table~\ref{tab:simple_approach_param}.}
    \label{fig:nirspec_lab}
\end{figure}

\begin{figure}
    \centering
    \includegraphics[width=0.95\linewidth]{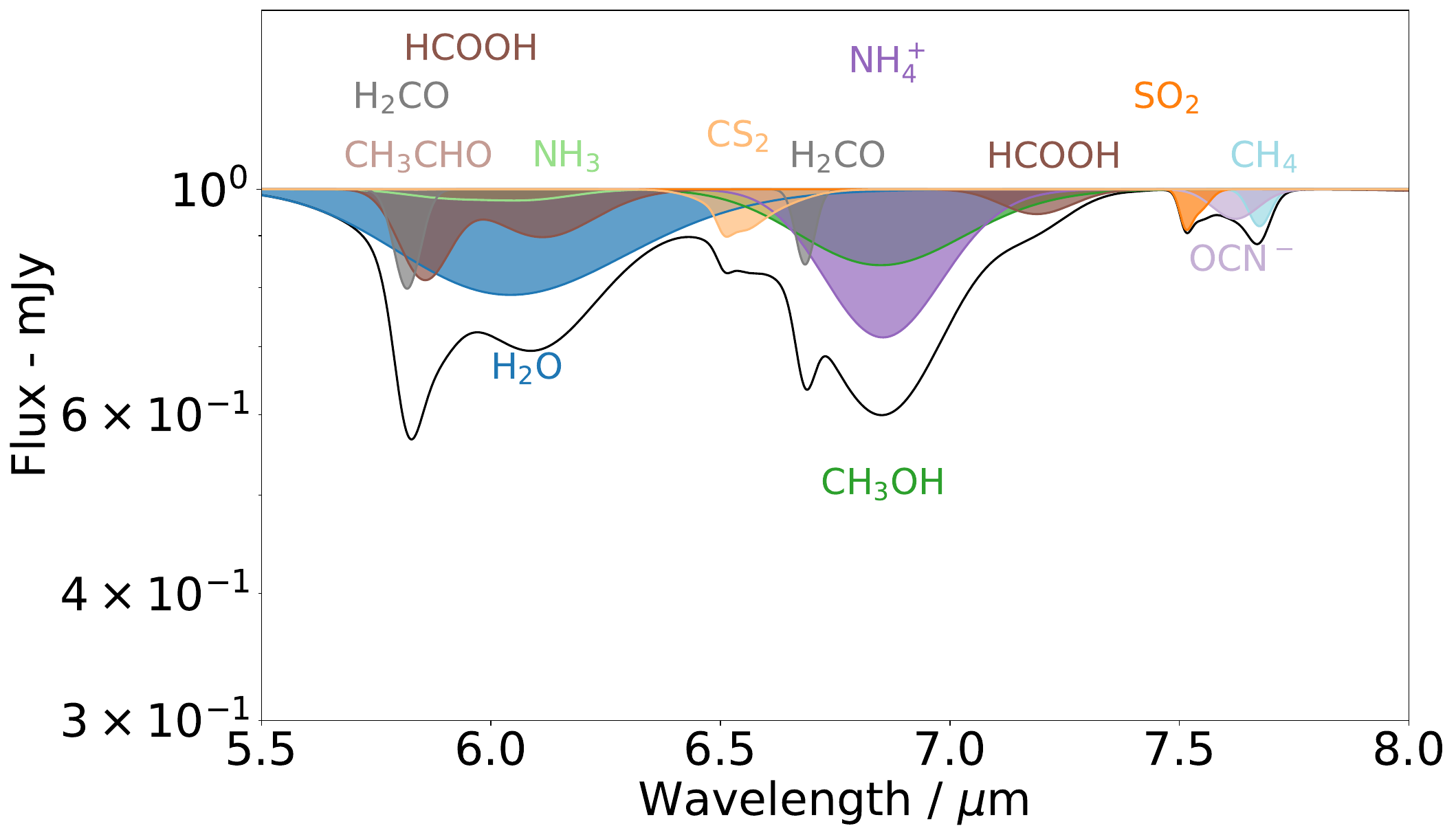}
    \caption{Similar to Fig.~\ref{fig:nirspec_lab} but for the different features and their contributions used in SynthIceSpec in the MIRI wavelength range from 6 to 8 $\mu$m. The overall spectrum is plotted in black. }
    \label{fig:miri8_lab}
\end{figure}

\begin{figure}
    \centering
    \includegraphics[width=0.95\linewidth]{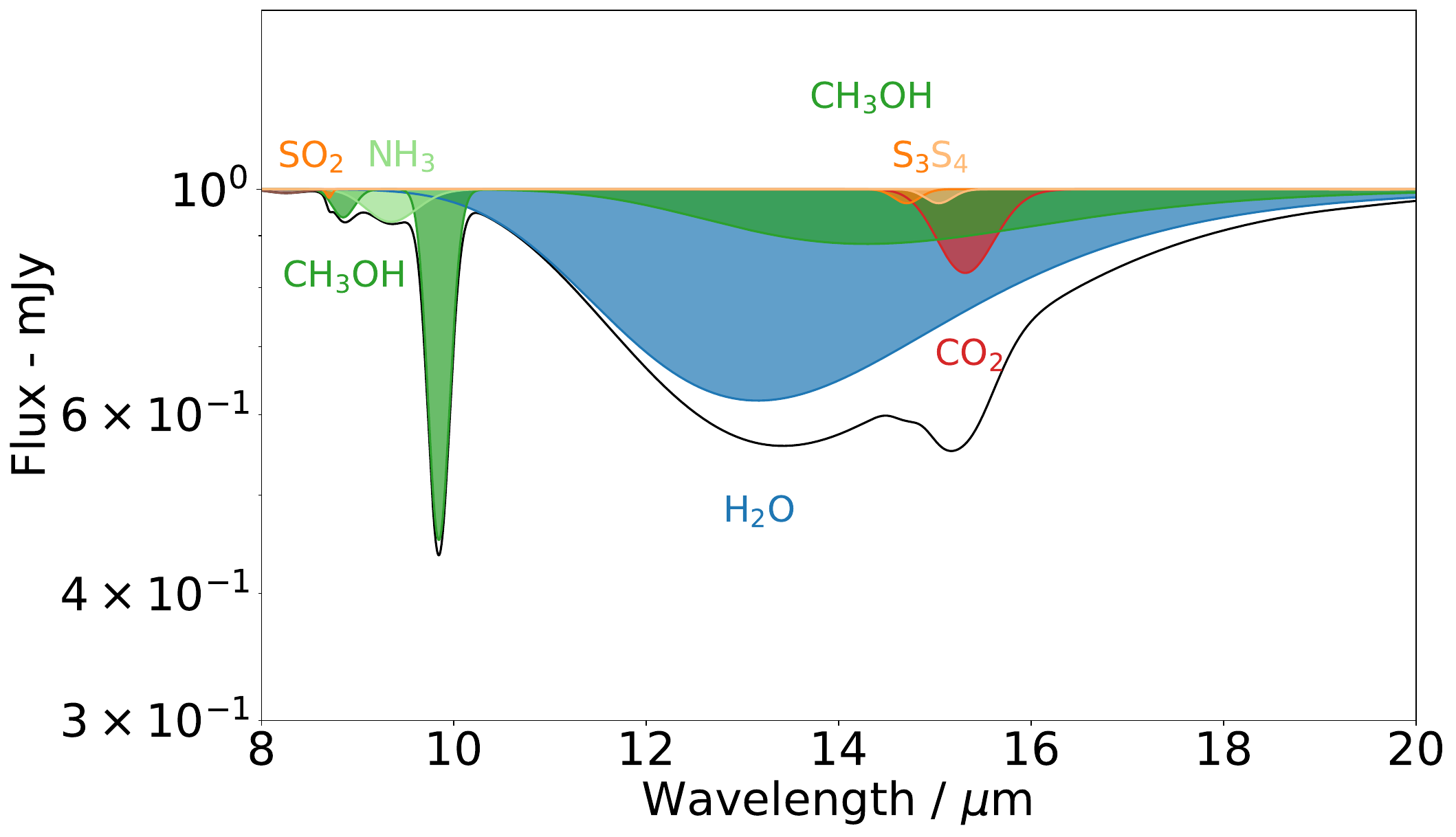}
    \caption{Same as Fig.~\ref{fig:miri8_lab} but on the MIRI wavelength range from 8 to 20 $\mu$m. The overall spectrum is plotted in black. }
    \label{fig:miri20_lab}
\end{figure}

In this Appendix, we present the different features available in SynthIceSpec, derived from laboratory spectra, in Figs.~\ref{fig:nirspec_lab}, \ref{fig:miri8_lab}, and \ref{fig:miri20_lab} for NIRSpec and MIRI (for two wavelengths ranges).

\section{Gas-grain model prediction for dense cloud environment}\label{appendixb}

\begin{figure*}[h!]
    \centering
    \includegraphics[width=0.99\linewidth]{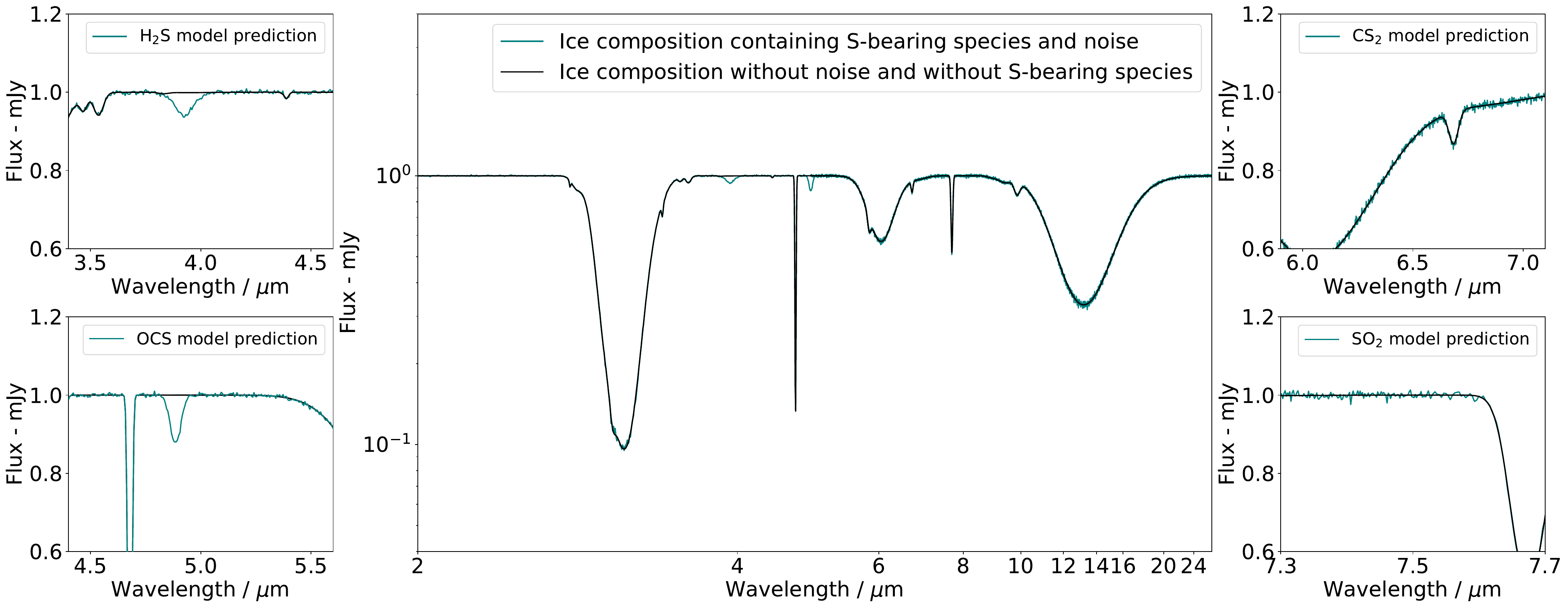}
    \includegraphics[width=0.99\linewidth]{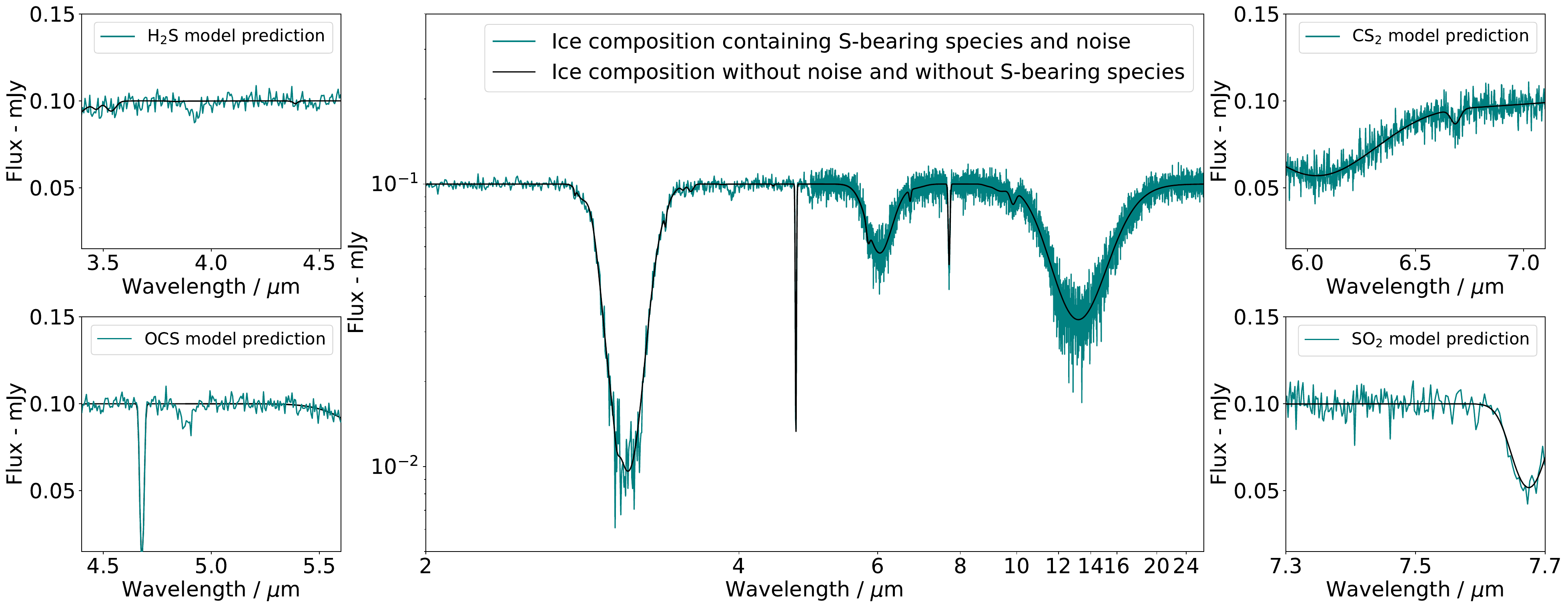}
    \includegraphics[width=0.99\linewidth]{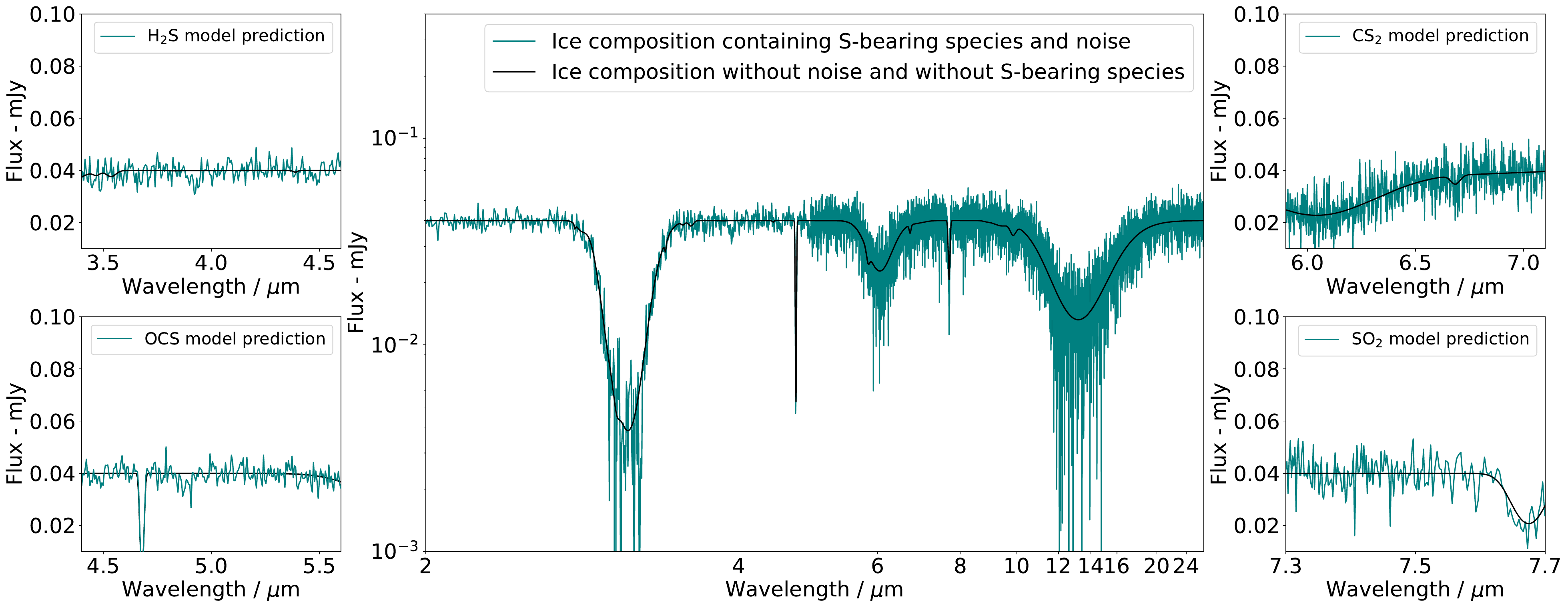}
    \caption{Threshold determination for the dense cloud environments for 3 different continua (1, 0.1 and 0.04 mJy) as predicted by Nautilus in TMC-1 physical conditions, the synthetic ice spectra in teal is the ice composition containing sulphur (and noise added), with in black the synthetic ice spectra without sulphur and without noise. Each panel is the 5$\sigma$ threshold for each species considered (on the left, top to bottom: H$_2$S, OCS; on the right, top to bottom: CS$_2$ , SO$_2$). } 
    \label{fig:bg_star_no_continuum_sis_nautilus}
\end{figure*}

As SynthIceSpec can take outputs of chemical gas-grain models, we used the ice predictions from \citet{navarro-almaida_gas_2020} to get an overview of the expected observations in the dense cloud phase. Based on TMC-1 observations, the authors used the Nautilus model \citep{ruaud_gas_2016,Wakelam_2024} to reproduce the gas-phase observations, including deuterated species and spin chemistry (at the gas and grain interface). 
The authors predict the H$_2$S ice abundance to be about one fifth of the sulphur cosmic abundance, with a value of $\sim$ 3.4 $\times$ 10$^{-6}$. We use the values reported in Table. 5 of their paper to reproduce the sulphur species.
H$_2$S is predicted to be the main sulphur bearing ice reservoir (2.5\% with respect to solid H$_2$O), followed by NS (1.3\%, not included in our synthetic spectra) and OCS (0.29\%). 
Other ice species abundances are derived from the models at the highest extinction point (``Minissale ice scheme" scenario) and implemented in our synthetic spectra. The column densities are listed in Table~\ref{tab:col_dens_nautilus}. The physical parameters of the model used are as follow: $n_{\rm H}$ = 6.4 $\times$ 10$^4$ cm$^3$, T = 10 K, t = 1 Myr, $\zeta_{\rm H_{2}}$ = 1.15 $\times$ 10$^{-16}$ s$^{-1}$, and $\chi = 5$ in Draine field units.
The synthetic spectra resulting from the model outputs is presented in Fig.~\ref{fig:bg_star_no_continuum_sis_nautilus}. 
With the predicted column densities, we verify if the detection would be possible in these conditions.
CS$_2$ and SO$_2$ column densities are too low to be detected at any continuum level. 
H$_2$S and OCS (2.5\% and 0.29\% respectively with respect to solid H$_2$O) are detected at more than 5$\sigma$ at a constant flux of 1 mJy. 
OCS is detected at a constant flux of 0.1 mJy, with a $\sim$5$\sigma$ level while H$_2$S is only at a $\sim$2$\sigma$ level for the same continuum.
Finally, only OCS is identified at a constant flux of 0.04 mJy with a $\sim$3$\sigma$ level.

\begin{table}[!h] 
\centering
\caption{Predicted column densities}
\begin{tabular}{|l|c|}
\hline
\multicolumn{1}{|c|}{Molecules} & \multicolumn{1}{c|}{Column density (molecules cm$^{-2}$)}                \\
\multicolumn{1}{|c|}{} & \multicolumn{1}{c|}{(\% H$_2$O)}                               \\
\hline
H$_2$O          & 9.3$\times$10$^{18}$          \\  
CO              & 1.6$\times$10$^{18}$ (26.0\%) \\   
CO$_2$          & 3.9$\times$10$^{13}$ ($<$ 0.1\%)  \\  
CH$_3$OH        & 2.2$\times$10$^{17}$ (3.7\%)   \\   
NH$_3$          & 2.9$\times$10$^{17}$ (4.9\%)   \\ 
NH$_4^+$        & 0 (0\%)   \\
CH$_4$          & 18.1$\times$10$^{17}$ (18.1\%)   \\ 
OCN$^-$         & 2.2$\times$10$^{13}$ ($<$0.1\%) \\
HCOOH           & 3.7$\times$10$^{11}$ ($<$0.1\%)    \\      
CH$_3$CHO       & 6.1$\times$10$^{11}$ ($<$0.1\%) \\
H$_2$CO         & 2.6$\times$10$^{17}$ (4.4\%)    \\
\hline
H$_2$S  &  \multicolumn{1}{c|}{1.5$\times$10$^{17}$ (2.5\%)}  \\
OCS     &  \multicolumn{1}{c|}{1.7$\times$10$^{16}$ (0.29\%)} \\
CS$_2$  &  \multicolumn{1}{c|}{0 (0\%)}  \\
SO$_2$  &  \multicolumn{1}{c|}{1.8$\times$10$^{12}$ ($<$0.1\%)} \\
\hline\end{tabular}
\tablefoot{Column densities predicted by Nautilus used in the synthetic spectra Fig.~\ref{fig:bg_star_no_continuum_sis_nautilus}}\label{tab:col_dens_nautilus}
\end{table}

\end{document}